\documentstyle[12pt]{article}
\font\tenrm=cmr10

\def\ga{\gamma}
\def\de{\delta}

\def\th{\theta}

\def\De{\Delta}

\def\fr#1#2{{{#1} \over {#2}}}
\def\prt{\partial}

\def\vev#1{\langle {#1}\rangle}

\def\frac#1#2{{\textstyle{{#1}\over {#2}}}}

\def\lsim{\mathrel{\rlap{\lower4pt\hbox{\hskip1pt$\sim$}}
    \raise1pt\hbox{$<$}}}
\def\gsim{\mathrel{\rlap{\lower4pt\hbox{\hskip1pt$\sim$}}
    \raise1pt\hbox{$>$}}}
\def\sqr#1#2{{\vcenter{\vbox{\hrule height.#2pt
         \hbox{\vrule width.#2pt height#1pt \kern#1pt
         \vrule width.#2pt}
         \hrule height.#2pt}}}}

\def\Re{\hbox{Re}\,}

\newcommand{\beq}{\begin{equation}}
\newcommand{\eeq}{\end{equation}}
\newcommand{\bea}{\begin{eqnarray}}
\newcommand{\eea}{\end{eqnarray}}
\newcommand{\rf}[1]{(\ref{#1})}

\renewenvironment{thebibliography}[1]
 { \rm
   \begin{list}{\arabic{enumi}.}
    {\usecounter{enumi} \setlength{\parsep}{0pt}
     \setlength{\itemsep}{3pt} \settowidth{\labelwidth}{#1.}
     \sloppy
    }}{\end{list}}

\def\theequation{\thesection.\arabic{equation}}
\def\@eqnnum{{\rm (\theequation)}}

\def\dif{{\rm d}}

\def\nue{{\nu_e}}
\def\num{{\nu_\mu}}
\def\nut{{\nu_\tau}}
\def\nueb{{\bar \nu_e}}
\def\numb{{\bar \nu_\mu}}
\def\nutb{{\bar \nu_\tau}}
\def\tgzero{\bar g_0}
\def\tgone{\bar g_1}
\def\tgtwo{\bar g_2}
\def\tgthr{\bar g_3}
\def\Ceko{C^1_e}
\def\Cekt{C^2_e}
\def\Cmko{C^1_\mu}
\def\Cmkt{C^2_\mu}

\def\CB{C}

\def\calB{{\cal B}}
\def\calC{{\cal C}}

\def\ptemp{ \fr{1}{64}
    \exp \left( \fr{E' - E}{2T} \right)
     \left( \fr{T}{E} \right)^2}

\def\sinsthw{\sin^2 \theta_w}

\def\Aek{ \left( { \fr{5 (a+b)+17}{3} } \right) }
\def\Amk{ \left( { \fr{5 (b+c)+17}{3} } \right) }
\def\Bek{ \left( { a+b } \right) }
\def\Bmk{ \left( { b+c } \right) }
\def\ergk{ \fr{E}{T} }
\def\fexp{ \exp  \left( { - \fr{E}{T} } \right) }
\def\ergkfac{ \ergk \left(
   { \fr{11}{12} \ergk - 1 } \right) \fexp }

\newcommand{\bref}[1]{(\ref{#1})}
\newcommand{\ct}[1]{\cite{#1}}
\def\lrb#1{\left( { #1 } \right)}
\def\EoTt{\left( {\frac{E}{T} ,t  } \right)}
\def\EpoTt{\left( {\frac{E'}{T} ,t } \right)}
\def\EoT{\left( {\frac{E}{T} } \right)}
\def\EoTEpoT{\left( { \frac{E}{T} , \frac{E'}{T} } \right)}
\def\ie{i.e.,\ }
\def\c2th{\cos 2\theta}
 \def\s2th{\sin2\theta}
\def\ss{\sin^2 2\theta}

\def\secttit#1{\vglue 0.6cm{\bf\large\noindent{#1}}\vglue 0.4cm}

\begin{document}
\titlepage

\begin{flushright}
{IUHET 298\\}
{MPI--PhT/95--25\\}
{CCNY--HEP--95/4\\}
{hep-ph/9507427\\}
{March 1995\\}
\end{flushright}
\vglue 1cm

\begin{center}
{
{\Large \bf
Neutrino Oscillations in the Early Universe
\\
with Nonequilibrium Neutrino Distributions
\\}
\vglue 1.0cm
{V. Alan Kosteleck\'y$^{a}$ and Stuart Samuel$^{b*}$\\}
\bigskip
{\it $^a$Physics Department\\}
{\it Indiana University\\}
{\it Bloomington, IN 47405, U.S.A.\\}
\vglue 0.3cm
\bigskip

{\it $^b$Max-Planck-Institut f\"ur Physik\\}
{\it Werner-Heisenberg-Institut\\}
{\it F\"ohringer Ring 6\\}
{\it 80805 Munich, Germany\\}

\vglue 0.8cm
}

\vglue 0.3cm

{\bf Abstract}
\end{center}
{\rightskip=3pc\leftskip=3pc\noindent
Around one second after the big bang,
neutrino decoupling and $e^+$-$e^-$ annihilation
distort the Fermi-Dirac spectrum of neutrino energies.
Assuming neutrinos have masses and can mix,
we compute the distortions using nonequilibrium thermodynamics
and the Boltzmann equation.
The flavor behavior of neutrinos is studied
during and following the generation of the distortion.
}

\vskip 1truein
\centerline{\it Accepted for publication in Physical Review D}
\vfill

\textwidth 6.5truein
\hrule width 5.cm
\vskip 0.3truecm
{\tenrm{
\noindent
$^*$Permanent address: Physics Department,
City College of New York,
New York, NY 10031, USA.\\
\hspace*{0.2cm}E-mail: samuel@scisun.sci.ccny.cuny.edu\\}}

\newpage

\baselineskip=20pt

{\bf\large\noindent I.\ Introduction}
\vglue 0.2cm
\setcounter{section}{1}
\setcounter{equation}{0}

Neutrinos play a significant role
both in cosmology
\cite{kt90a}
and in particle physics
\cite{bv92}.
In cosmology,
during the radiation-dominated period
in the early-Universe
from one second after the big bang
to around twenty-thousand years,
neutrinos are almost as important as photons
in driving the expansion of the Universe.
Furthermore,
nucleosynthesis and the ensuing abundances of light elements
are strongly influenced by the number and nature of neutrinos.
In particle physics,
it is hoped that current oscillation experiments
coupled with solar and atmospheric neutrino observations
will lead to insight beyond the standard model.
The basic idea is that interference among neutrino flavors
provides a sensitive experimental probe
for neutrino masses and mixings.

It is natural to ask whether neutrino oscillations
affect the physics of the early Universe.
Different aspects of this question have been addressed by
a number of authors
\ct{d81a,s87a,l,n,s,ekm91a,ekt92,ssf93a,kps93a,ks93a,ks94a,mt94a}.
When the temperature $T$ of the Universe
is several MeV and higher,
interactions keep neutrinos in thermal contact
with electrons, positrons and the primodial plasma
and so oscillations cannot occur.
If thermal equilibrium were maintained for all times,
then the numbers of electron, muon, and tau neutrinos
would be equal.
This would preclude flavor oscillations because,
for instance,
for each $\nu_e$ that converts into a $\nu_\mu$
there would be a $\nu_\mu$ converting into a $\nu_e$.

However,
around one second when $T$ is about $1$ MeV,
nonequilibrium distributions develop.
This effect is due to the coincidence
of thermal-neutrino decoupling and $e^+$-$e^-$ annihilation.
As some $e^+$-$e^-$ pairs annihilate,
they reheat photons and other electrons and positrons.
The latter in turn can interact with neutrinos
via the weak interactions,
thereby slightly reheating the neutrinos.
This reheating is both flavor and energy dependent
because the neutrinos are decoupling at this time.
Due to $W^\pm$ exchanges,
electron neutrinos are heated somewhat more
than muon and tau neutrinos.
This leads to an excess of electron neutrinos
over muon and tau neutrinos.
Furthermore, higher-energy neutrinos interact
more strongly than lower-energy neutrinos
so that a relative excess of higher-energy neutrinos arises.
The appearance of a flavor excess means that
neutrino oscillations can occur.
This potentially could affect nucleosynthesis
\cite{l}.

The process generating the distorted distributions
is sensitive to the timing of events in the early Universe.
If the neutrinos had decoupled well before
$e^+$-$e^-$ annihilation,
then they would have maintained Fermi-Dirac distributions.
Their energies and momenta
would simply have been redshifted
by the inverse scale factor $R^{-1} (t)$
of the Friedmann-Robertson-Walker (FRW) cosmology.
For ultra-relativistic particles,
such redshifts maintain standard statistical distributions.
If instead the neutrinos had decoupled well after
$e^+$-$e^-$ annihilation,
then they would have remained in thermal equilibrium
during the heating process.
When they decoupled later,
they would have again maintained
a redshifted Fermi-Dirac distribution.
In either case,
no flavor asymmetry would have arisen
and so no oscillations would have occurred.

The analysis of nonequilibrium neutrino distributions
is not an easy undertaking,
even in the absence of neutrino mixing.
This problem has recently been treated
in two different approaches based on the Boltzmann equation.
A detailed study allowing for contributions
from all tree-level scattering amplitudes,
performing exactly many phase space integrations,
and using numerical methods to solve the Boltzmann equation
is presented in
ref.\ \ct{dt92a}.
Estimates of the effects
and relatively simple approximate analytical formulae
for the distortion of distributions are provided
in ref.\ \ct{df}.

When neutrinos mix,
the problem of analyzing nonequilibrium neutrino distributions
becomes significantly more complicated.
The complications arise not only
from the simultaneous occurrence of production and oscillations
but also from indirect effects
modifying vacuum-oscillation behavior.
These effects arise from neutrino interactions
with the background gas of electrons, positrons,
and other neutrinos.
In particular,
neutrino-neutrino interactions
can strongly affect oscillations
because neutrinos in the early Universe form a dense gas
and because the effects are nonlinear.
These interactions enter as flavor off-diagonal
as well as flavor diagonal terms
in the effective hamiltonian
\ct{pantaleone}.
The ensuing complications can be handled
using Hartree-Fock-like
or density-matrix formalisms
\ct{sr93a,samuel93a}.

In previous papers
\ct{kps93a,ks93a,ks94a},
we have analyzed neutrino-flavor properties in the early Universe
assuming nonequilibrium neutrino distributions
based on the approximate analytical formulae
of ref.\ \ct{df}.
These analyses therefore disregard
the role of neutrino oscillations during
the generation of the nonequilibrium distributions,
although various tests suggest
that our qualitative results are correct.

In the current work,
we obtain the equations that describe
the nonequilibrium distortions of neutrino distributions
when mixing is present,
and we present numerical solutions
of the ensuing neutrino behavior.
The remainder of this introduction
provides a guide to the structure of the paper.

For simplicity,
we assume that mixing occurs only between two neutrinos,
taken to be $\nu_e$ and $\nu_\mu$,
and that the third neutrino,
taken as $\nu_\tau$,
does not participate in the oscillations.
The analysis of nonequilibrium statistics
in ref.\ \ct{dt92a}
was performed under the assumption that
muon and tau neutrinos behave identically.
However,
with neutrino mixing,
the tau neutrino cannot be disregarded
because it participates
in the generation of thermal distortions
via $e^+$-$e^-$ annihilations.
The equations obtained in ref.\ \ct{dt92a}
must therefore be generalized.
This is accomplished in Sect.\ II.

Section III reviews our formalism
for dense-neutrino oscillations in the early Universe,
while Sect.\ IV summarizes
the neutrino-oscillation behavior uncovered
in refs.\ \ct{kps93a,ks94a}.
This summary is used in the discussion of
our new results in Sections VIII and IX.
Section V combines the results of
Sects.\ II and III to obtain equations
that govern the production
of nonequilibrium neutrino distributions
in the presence of mixing.

The deviations
from standard thermal distributions
are generated
from about $0.1$ seconds to about $1.5$ seconds.
We call this the {\it production phase}
because the flavor-dependent neutrino excesses
are produced during this interval.
By {\it production profile},
we mean the nonequilibrium distortions
which arise at the end of the production phase.
The term {\it pure production}
is reserved for production
without neutrino mixing.
After about $1.5$ seconds,
neutrino decoupling is sufficiently strong
that few further distortions
in statistical distributions are generated.
We call the period after about $1.5$ seconds
the {\it oscillation phase}.

One of our goals is to understand
neutrino-flavor variation during
the production phase.
This requires computer simulations
because the equations are nonlinear and relatively complicated.
Issues concerning numerical methodology
are discussed in Sect.\ VI.
Some improvements are made over the previous work
of ref.\ \ct{dt92a}
even for the pure-production case.
For this reason,
we have repeated the analysis
of the pure-production case in Sect.\ VII.

Another goal is to obtain the production profile
in the presence of neutrino mixing.
Production profiles for various neutrino mixing angles
and mass differences are presented in Sect.\ VIII.
This section also discusses
the general behavior of neutrino oscillations
and related properties.

Section IX
analyzes the oscillation phase.
The main purpose is to check the results of
refs.\ \ct{kps93a,ks94a}.
The qualitative flavor properties
described in \ct{kps93a,ks94a} are confirmed.
Small numerical differences are uncovered,
however.
These can be attributed to the use of the
approximate analytical formulae for the production profile.
Finally, we summarize in Sect.\ X.

An important improvement of the current work
over refs.\ \ct{kps93a,ks94a}
is that the overall normalization of distortions
is incorporated.
The earlier works explicitly avoided this issue
by considering ratios that eliminated the normalization factor.
Our current simulations are now relatively
accurate in absolute terms.
At all times,
the numbers of excess electron, muon, and tau neutrinos
per cubic volume are specified to within about $25 \%$.
The uncertainties are dominated by systematic effects,
discussed in Sect.\ VI.

Throughout this paper,
we work in units with $k = \hbar = c = 1$.
The values of various parameters
used in our current work,
e.g., the baryon-to-photon ratio $\eta$,
coincide with those of ref.\ \ct{ks94a}.
It is assumed that
the chemical potentials for neutrinos are zero,
so that the total number of neutrinos and antineutrinos
is the same.

\bigskip
\vglue 0.2cm
{\bf\large\noindent
II.\ Pure Production for Three Flavors}
\vglue 0.2cm
\setcounter{section}{2}
\setcounter{equation}{0}

Under the assumption that
neutrinos are massless or do not mix,
the distortion of the neutrino Fermi-Dirac distribution
induced by the combination of neutrino thermal decoupling
and $e^+$-$e^-$ annihilation has been determined in
ref.\ \ct{dt92a}
using the Boltzmann equation in an expanding FRW cosmology.
For this situation,
the distortions in the muon-neutrino
and tau-neutrino distributions are equal,
as are the distortions for antineutrinos and neutrinos
of the same flavor.
However,
in the presence of mixing and neutrino oscillations,
these equalities do not hold,
and it is necessary to generalize the results of
ref.\ \ct{dt92a}.
The purpose of this section is to obtain
the generalized equations.

We begin by listing four of the more important approximations
made in the analysis.
First,
at the energies and temperatures of interest,
the difference between Fermi-Dirac
and Maxwell-Boltzmann statistics is unimportant
because there is little fermion degeneracy.
For calculational purposes
it is convenient to use Maxwell-Boltzmann statistics.
A second approximation is the restriction to dominant
weak-interaction effects,
of order $G_F^2$.
This is an excellent approximation
during the time interval of interest in the early Universe.

A third approximation concerns the value of
\beq
\delta (t) \equiv T_\gamma/T - 1
\quad ,
\label{2p1}
\eeq
where $T_\ga$ is the photon temperature and
$T$ is the neutrino temperature.
The quantity $\de(t)$ is a measure of
the photon-neutrino temperature difference $T_\gamma - T$,
to which the distortions
of the neutrino distributions are proportional.
An exact computation of $\delta (t)$ is infeasible.
However,
it may be approximated by $\delta_0 (t)$,
where $\delta_0 (t)$ is obtained assuming that neutrinos
are always thermally decoupled
and that they do not share
in the heat released by $e^+$-$e^-$ annihilations.
While this approximation is poor for early times
when the temperature is above $\sim 5$ MeV,
the corresponding values of $\delta (t)$ are relatively small,
and so the error introduced in the distortion is minimal.
A correction to $\delta_0 (t)$ partially compensating
for the approximation is given in
ref.\ \ct{dt92a}.

A fourth approximation
is to set the electron mass $m_e$ to zero.
This is particularly convenient
when computing scattering amplitudes.
The approximation is reasonable for $T \gsim 0.5$ MeV.
It does not introduce a large error for $T < 0.5$ MeV
because most of the production occurs
at temperatures above $0.5$ MeV.

We next introduce the neutrino distortions.
When neutrinos are in thermal equilibrium,
their energy distribution is governed by
the Maxwell-Boltzmann factor $f_0 (E)$ given by
\beq
 f_0 (E) = \exp \lrb{ -\fr{E}{T} }
\quad ,
\label{2p2}
\eeq
where $E$ is the neutrino energy.
As $T$ drops below $\sim 5$ MeV,
neutrinos decouple thermally.
The distribution $f_{\nu_f} \lrb{ E, t}$
for each neutrino $\nu_f$ of flavor $f$,
${\nu_f} = \nue$, $\num$, $\nut$, $\nueb$, $\numb$ or $\nutb$,
then deviates from the Maxwell-Boltzmann one.
We write
\beq
  f_{\nu_f} \lrb{ E, t} =
    f_0 (E) + \Delta_{\nu_f} \lrb{ E, t}
\quad .
\label{2p3}
\eeq
This equation defines the distortions
$\Delta_{\nu_f} \lrb{ E, t}$.

For early times $t$ when the system is hot,
$\Delta_{\nu_f} (E,t) = 0$, $f_{\nu_f}(E) =  f_0 (E)$
and there is no deviation from standard statistics.
For later times,
$\Delta_{\nu_f} \lrb{E,t}$ is nonzero
but small compared to $f_0 (E)$
because the neutrinos are weakly interacting
and are only mildly sensitive
to the reheating from $e^+$-$e^-$ annihilation.
Thus, $\Delta_{\nu_f} \lrb{E,t}$ can be treated
as a perturbation.
Since electron neutrinos are somewhat more sensitive
to $e^+$-$e^-$ annihilation than muon or tau neutrinos,
$\Delta_\nue$ is larger than $\Delta_\num$ or $\Delta_\nut$.

The procedure for obtaining equations
for the $\Delta_{\nu_f}$ is given in
ref.\ \ct{dt92a}.
We therefore restrict ourselves here
to outlining our derivation of the generalized equations
for the distortions.
Note that, throughout the calculations,
each neutrino momentum $p$ and energy $E$
can be equated.
This is an excellent approximation
since the neutrinos are ultra-relativistic
at the temperatures of interest.
Also,
in working with the Boltzmann equation
for $f_{\nu_f} (E)$ in an expanding Universe,
it is convenient to use the variable $E/T$
whenever possible since, for relativistic particles,
it does not redshift as the Universe expands.
The FRW scale factor $R (t)$ evolves to maintain
$T(t) R(t)$ constant to a good approximation
because neutrinos are decoupling and
do not participate heavily in the $e^+$-$e^-$ reheating.
In contrast,
the photon temperature $T_\gamma$ rises relative
to $T$ during $e^+$-$e^-$ annihilations.

The squared matrix elements
of neutrino-scattering processes
appear in the Boltzmann equation.
There are $42$ different basic processes:
$$
     \nu_a + \bar\nu_a \to        e^+ + e^-
\quad , \quad
     \nu_a + \bar\nu_a \to      \nu_b + \bar\nu_b
\quad ,
$$
$$
     \nu_a + e^-       \to      \nu_a + e^-
\quad , \quad \quad
 \bar\nu_a + e^-       \to  \bar\nu_a + e^-
\quad ,
$$
$$
     \nu_a + e^+       \to      \nu_a + e^+
\quad , \quad \quad
 \bar\nu_a + e^+       \to  \bar\nu_a + e^+
\quad ,
$$
$$
     \nu_a +     \nu_b \to      \nu_a +     \nu_b
\quad , \quad \quad
 \bar\nu_a + \bar\nu_b \to  \bar\nu_a + \bar\nu_b
\quad ,
$$
\beq
     \nu_a + \bar\nu_b \to      \nu_a + \bar\nu_b
\quad , \quad a \ne b
\quad ,
\label{2p4}
\eeq
where $a$ and $b$ stand for $e$, $\mu$, or $\tau$.
In the Boltzmann equation,
an initial neutrino is distinguished.
This leads to $24$ additional cases.
Thus, there are a total of $66$
distinguished-neutrino processes.
For each particular neutrino or antineutrino,
there are $11$ cases.

In the calculation,
hard scattering of neutrinos off nucleons
can be neglected because the nucleon density is small.
Also,
nucleons do not affect neutrino oscillations through
forward scattering because electron and muon neutrinos
are scattered in the same way.
As noted above,
processes involving electrons and positrons
tend to reheat the neutrinos as $e^+$ and $e^-$ annihilate,
and electron neutrinos participate more in the reheating
due to charge-current interactions from $W^\pm$ exchange.
In contrast,
pure neutrino and antineutrino processes
tend to equilibriate the distributions.

The electron and positron Maxwell-Boltzmann distributions
are
\beq
  f_{e^\pm} \lrb{ E_e, t} =
     \exp \lrb{ -\fr{E_e}{T_\gamma} }
\quad .
\label{2p6}
\eeq
Since the electron mass is set to zero,
$E_e = p_e$.
Expressing Eq.\ \bref{2p6}
in terms of the neutrino temperature $T$
and the fractional temperature difference $\delta (t)$
gives
\beq
  f_{e^\pm} \lrb{ p_e, t} =
     f_0 (p_e) \lrb{ 1 + \fr{p_e}{T} \delta (t) + \dots }
\quad .
\label{2p7}
\eeq

The Boltzmann equation itself
and the squared matrix elements
for all the processes in Eq.\ \bref{2p4}
are given in ref.\ \ct{dt92a}.
To obtain equations for the distortions,
Eqs.\ \bref{2p3} and \bref{2p7}
and the squared matrix elements
are substituted into the Boltzmann equation,
and an expansion is performed in powers of
$\Delta_{\nu_a}$, $\Delta_{\bar \nu_a}$ and $\delta (t)$.
The term linear in these quantities provides
differential equations for the $\Delta_{\nu_f}$.
For calculational purposes,
it is useful to consider the distortions $\Delta_{\nu_f}$
and other quantities as functions of ${E}/{T}$ and $t$,
rather than $E$ and $t$.

The resulting differential equations can be simplified
by performing many of the phase space integrations.
After some calculation,
we obtain for the six distortions $\Delta_{\nu_f}$
a set of six linear coupled differential equations,
given by
$$
{{ d \Delta_{\nu_f}} \over {dt}} \EoTt =
  { {4G_F^2 T^5} \over {\pi^3} }
  \left( {
       - A_{\nu_f} \EoT
    \Delta_{\nu_f} \EoTt
         } \right.
\qquad\qquad
\qquad\qquad
\qquad\qquad
$$
\beq
  + B_{\nu_f} \EoT \delta \left( t \right) +
   \sum\limits_{\nu_{f'}}
   \int\limits_0^\infty  {{{ \dif E'} \over T}}
  \left. {
   C_{\nu_f\nu_{f'}} \EoTEpoT
   \Delta_{\nu_{f'}} \EpoTt
         } \right)
\quad ,
\label{2p8}
\eeq
where the $A$, $B$, and $C$ coefficients are determined.
They are presented in Appendix A.

Since $A_{\nu_f} > 0$,
the $A$ terms induce damping.
The $B$ terms induce reheating of neutrinos,
arising from interactions with positrons and electrons.
Note that our conventions for the $A$ and $B$
coefficients differ from ref.\ \ct{dt92a}
by a factor of $E/\lrb{T \pi^3 }$,
included here as part of the overall normalization constant
in Eq.\ \bref{2p8}.
The $C$ terms
represent the coupling of neutrinos of different energies.
This is the origin of the integration
over the final state neutrino energy $E'$.
Note that all the coefficients are independent of $t$,
since they are functions only of $E/T$ and $E'/T$.

We remark that Eqs.\ \bref{2p8}
correctly reduce to the corresponding equations%
\footnote{Note that the factor $(c+8)$ in Eq.\ (2.11d)
of ref.\ \ct{dt92a} should be $(c+7)$,
and the factor $(b+c+14)$ in Eq.\ (2.14b)
should be $(b+c+13)$.}
of ref.\ \ct{dt92a} in the limits
\beq
  \Delta_\nueb \to \Delta_\nue
\ , \quad \quad
  \Delta_\nut  \to \Delta_\num
\ , \quad \quad
  \Delta_\numb \to \Delta_\num
\ , \quad \quad
  \Delta_\nutb \to \Delta_\num
\quad .
\label{2p13}
\eeq

\bigskip
\vglue 0.2cm
{\bf\large\noindent
III.\ Pure Oscillations for Two Flavors}
\vglue 0.2cm
\setcounter{section}{3}
\setcounter{equation}{0}

In this section,
we summarize the formalism of our earlier work
for treating neutrino oscillations
in the early Universe.
We consider oscillations between two flavors
$\nue$ and $\num$.

The most convenient formulation uses
the three-component vectors
$$
  \vec v (E,t) \equiv
 \left( {\nu_e^{\dag} \nu_e    -
       \nu_\mu^{\dag} \nu_\mu ,
 2 Re ( {\nu_e^{\dag} \nu_\mu  }),
 2 Im ( {\nu_e^{\dag} \nu_\mu  } )}
 \right)
\quad ,
$$
\beq
 \vec w (E,t) \equiv
 \left( {\bar \nu_e^{\dag} \bar\nu_e    -
        \bar\nu_\mu^{\dag} \bar\nu_\mu  ,
 2 Re ( {\bar \nu_e^{\dag} \bar\nu_\mu  } ),
 2 Im ( {\bar \nu_e^{\dag} \bar\nu_\mu  } )}
 \right)
\quad ,
\label{3p1}
\eeq
where $\nu_f$ and $\bar \nu_f$
are fields for neutrinos and antineutrinos of flavor $f$,
respectively.
We use the normalization convention that
$\nu_f^{\dag}\nu_f (E)$ is the number of neutrinos
of flavor $f$ with energy $E$
in a comoving volume of volume $a^3$
per unit $E/T$.
The scaling factor $a$ can be chosen arbitrarily.
It increases with the redshift scale
$R(t)$ as the Universe expands.

The oscillations equations resemble
the motion of a particle in a magnetic field:
\beq
 {{\prt \vec v (E)} \over {\prt t}} =
        \vec v (E) \times \vec B_v (E)
\ , \quad \quad
 {{\prt \vec w (E)} \over {\prt t}} =
        \vec w (E) \times \vec B_w (E)
\quad ,
\label{3p2}
\eeq
where the effective magnetic fields are
$$
\vec B_v (E, t) = {{\vec \Delta } \over {2E}} -
    {\vec V_{\nu \nu}} -
   \left( {{V_{CP^+} (E)} + {V_{CP^-}} } \right) \hat e_1
 \quad ,
$$
\beq
\vec B_w (E, t) = {{\vec \Delta } \over {2E}} +
  {\vec V^{*}_{\nu \nu}} -
   \left( {V_{CP^+} (E)} - {V_{CP^-}} \right) \hat e_1
\quad .
\label{3p3}
\eeq
The next few paragraphs explain the terms
in Eq.\ \bref{3p3}.

Vacuum oscillations are produced by
the term $\vec \Delta /(2E) $,
where
\beq
  {\vec \Delta } =
        \Delta \left( {\cos {2\theta },
                      -\sin {2\theta },0} \right)
\quad .
\label{3p4}
\eeq
Here, $\theta$ is the vacuum mixing angle
and $\Delta$ is the mass-squared difference
of vacuum-mass-eigenstate neutrinos:
$\Delta = m_2^2 - m_1^2$.
For pure vacuum oscillations,
neutrinos are in mass eigenstates
when $\vec v$ and $\vec \Delta$ are aligned.

The interaction of neutrinos with background
electrons and positrons is governed by two potentials,
$V_{CP^+} $ and $V_{CP^-}$, where
$V_{CP^+} $ is CP-conserving
and $V_{CP^-}$ is CP-violating:
$$
 {{V_{CP^+} (E , t )} } = - {{2 \sqrt 2 G_F E}
  \left( {\rho_{e^-} + p_{e^-} +
          \rho_{e^+} + p_{e^+}}
  \right)/{M_W^2}}
\quad ,
$$
\beq
 {V_{CP^-}} (t) = \sqrt 2G_F
  \left( {n_{e^-} - n_{e^+}} \right)
\quad ,
\label{3p5}
\eeq
where $n_f$, $\rho_f$ and $p_f$ are respectively
the number density,
the energy density and the pressure
of the charged lepton $f$.
As the Universe expands,
these quantities decrease.
The effective magnetic fields
associated with $V_{CP^+} $ and $V_{CP^-}$
are aligned along the $1$-axis
fixed as
$\hat e_1 = \left( {1,0,0} \right)$.
When neutrinos are in pure flavor eigenstates,
the $\vec v$ are aligned with $\hat e_1 $.

Finally, $\vec V_{\nu \nu}$ provides
the neutrino-neutrino interactions.
It is given by
\beq
  {{\vec V_{\nu \nu}} } (t) =
   \fr{\sqrt 2 G_F}{a^3} \left( {\left\langle {\vec v}
   \right\rangle -
   \left\langle {\vec w^*} \right\rangle} \right)
\quad ,
\label{3p6}
\eeq
where
\beq
  \left\langle {\vec v} (t) \right\rangle =
   \int\limits_0^\infty \fr{\dif E}{T} {\vec v (E , t) }
\ , \quad \quad
  \left\langle {\vec w} (t) \right\rangle =
   \int\limits_0^\infty \fr{\dif E}{T} {\vec w (E , t) }
\quad .
\label{3p7}
\eeq
The asterisk
in Eqs.\ \bref{3p3} and  \bref{3p6}
indicates that the sign
of the third component of a vector is reversed,
e.g.,
$\lrb{w_1 , w_2 , w_ 3}^* =  \lrb{w_1 , w_2 , - w_ 3}$.
Since $\vec V_{\nu \nu}$
renders the oscillation equations
in Eq.\ \bref{3p2} nonlinear,
we often refer to it
as the {\it nonlinear term}.

\bigskip
{\bf\large\noindent
IV.\ Summary of Results for
Pure Neutrino Oscillations}
\vglue 0.2cm
\setcounter{section}{4}
\setcounter{equation}{0}

For purposes of comparison with the
new results given in the present paper,
this section provides
a summary of the main features of neutrino behavior
in the early Universe described in
refs.\ \ct{kps93a,ks94a}.
Numerical solution of
Eq.\ \bref{3p2} for $\Delta > 0$ and $0 < \theta < \pi/4$
reveals that neutrino oscillations exhibit a number
of effects,
all attributable
to the nonlinear neutrino-neutrino interaction term.

One feature of the neutrino behavior
is its remarkable smoothness.
As the Universe expands,
the four interaction terms in Eq.\ \bref{3p3} change.
In the absence of the nonlinear term,
the changes induce temporary oscillatory behavior.
However,
when the nonlinear term is present,
no oscillations are observed.

Another feature of the behavior
concerns decoherence.
Without the nonlinear term,
decoherence is a collective effect
due to the different oscillation times of individual neutrinos.
Individual neutrinos
nonetheless undergo oscillatory behavior.
In contrast,
when the nonlinear term is present,
individual neutrino vectors
are approximately aligned
and so do not exhibit independent oscillatory behavior.

A third feature
of the nonlinear system is
that individual neutrinos maintain themselves in
configurations that are
approximate nonlinear mass eigenstates (ANME).
An instantaneous nonlinear mass eigenstate
diagonalizes the hamiltonian at a given time.
This occurs for an individual neutrino
when its vector and magnetic field
point in the same direction,
so the right-hand side of Eq.\ \bref{3p2} vanishes
and the change in flavor content is momentarily zero.
As the Universe expands,
magnetic fields evolve.
The ANME property
means that a neutrino vector tends to follow
changes in the associated magnetic field.
It can be shown that alignment is
a consequence of the ANME property
and the dominance of the neutrino-neutrino interaction.

A fourth feature of neutrino behavior is CP suppression.
Although CP-violating interactions are present,
the buildup of asymmetry
between neutrinos and antineutrinos
is several magnitudes smaller than in the absence
of the nonlinear neutrino-neutrino interactions.
The CP-suppression mechanism can also be understood
as a consequence of the ANME property.
It involves a partial cancellation
between the CP-asymmetric electron-neutrino
and neutrino-neutrino interactions.
This cancellation
can be used to develop an analytical approximation scheme
for the average and individual neutrino vectors.
The scheme reproduces
individual neutrino vectors to a few percent
and the average vector to better than one percent.

The neutrino behavior also exhibits planarity,
i.e.,
the third components of neutrino vectors
are much smaller than the first and second components.
In a linear system for which decoherence has set in,
planarity holds for the average vector
but not for individual neutrino vectors.
Since instantaneous nonlinear mass eigenstates
have zero third components,
the ANME property implies that
the third components of individual vectors are also small
for the full nonlinear system.

Since the third components are small,
it is tempting to conclude that
they are unimportant.
This is false.
Disregarding them,
which is equivalent to disregarding
the imaginary part of the off-diagonal flavor term,
is a poor approximation.
While the first two components
cancel to a large extent in the nonlinear term
of Eq.\ \bref{3p6},
the third components tend to add
because of the asterisk operation.
Numerically,
we find the third component
of $\vec V_{\nu\nu}$
is of the same order of magnitude
as the first two components.

Another feature of the neutrino behavior
can be traced directly to the third component
of $\vec V_{\nu\nu}$.
When $\Delta < 10^{-9}$ eV$^2$,
ANME configurations initially point approximately
along the $1$-axis.
As the Universe expands and energies redshift,
the vacuum term grows,
becoming dominant some tens of seconds later.
The magnetic fields then point toward $\vec \Delta$,
so ANME configurations are close to vacuum-mass eigenstates.
Surprisingly,
however,
the neutrino vectors rotate toward
vacuum-mass eigenstates earlier than expected.
This effect,
called precocious rotation,
arises because a negative contribution develops
to the third component of $\vec V_{\nu\nu}$,
which causes a rotation of the neutrino vectors
in the $1$--$2$ plane.

In passing,
we mention two other effects
involving the imaginary part of the off-diagonal flavor term,
arising in the $\Delta < 0$ parameter region
\ct{ks93a}.
First,
there exists a flavor-conversion mechanism
that is completely different from the MSW effect
\ct{msw}.
It is efficient and occurs even for very small mixing angles.
The second effect,
called self-maintained coherence,
is a collective mode of the nonlinear system
in which a large fraction of neutrinos oscillate in unison.
The effect in the pure neutrino gas
\ct{samuel93a}
differs from that in the mixed neutrino-antineutrino gas
\ct{ks94a,ks94b},
where the third component of $\vec V_{\nu\nu}$
plays an important role.

\bigskip
\vglue 0.2cm
{\bf\large\noindent
V.\ Equations for Combined Production and Oscillations}
\vglue 0.2cm
\setcounter{section}{5}
\setcounter{equation}{0}

If neutrino masses and mixings are nonzero,
neutrino oscillations begin
as distortions develop in the neutrino distributions.
In principle,
the oscillations could significantly affect
the flavor content when the photon reheating and
neutrino decoupling are complete.
For large mixing angle,
for instance,
any excess electron neutrinos produced
could oscillate into muon neutrinos,
potentially producing a muon-neutrino excess
that in the absence of oscillations would not arise.
Furthermore,
any change in flavor content due to oscillations
affects the rates of the scattering processes
in the Boltzmann equation.
The goal of this section is to obtain equations
governing the production of nonequilibrium distributions
when neutrino mixing is present.

During an infinitestimal time interval,
the change in the neutrino distributions
is due to the change created by the Boltzmann equation
plus the change due to neutrino oscillations.
Hence, it suffices to add the contributions
from pure production and from pure oscillations.
This involves combining the equations
in Sects.\ II and III.
However,
the formulae in Sect.\ II
are expressed in a different basis from
those in Sect.\ III.
For numerical purposes,
we elect to use the vector-type variables
of Sect.\ III.
We therefore begin this section
by rewriting the pure-production equations
of Sect.\ II.

The first step in this process is to relate
the distortions $\Delta_{\nu_f}$
to the density-like variables $\nu_f^{\dag} \nu_f $.
The connection is given by
\beq
  \Delta_{\nu_f} \lrb{ E, t}
  {\left( { a T } \right)^3 \over {2\pi^2}}
  \left( {{E \over T}} \right)^2
  = \nu_f^{\dag}\nu_f \left( {E,t} \right)
\quad .
\label{5p1}
\eeq
Note that the factor
$
  {{E^2 a^3 T} / \lrb{2\pi^2} }
$
is independent of time.
Multiplying both sides of
Eq.\ \bref{2p8} by this factor
leads to the reformulation of the pure-production equations as
$$
  {{\prt \nu_f^{\dag}\nu_f} \over {\prt t}} \left( {E,t} \right) =
  {{4G_F^2T^5} \over {\pi^3}}
 \left( {
                - A_{\nu_f} \EoT
         \nu_f^{\dag}\nu_f \left( {E,t} \right)
        } \right.
\qquad\qquad
\qquad\qquad
\qquad\qquad
$$
\beq
 + \tilde B_{\nu_f} \EoT
           \delta \left( t \right) +
   \sum\limits_{\nu_{f'}}
   \int\limits_0^\infty  {{{ \dif E'} \over T}}
\left. {
   \tilde C_{\nu_f\nu_{f'}} \EoTEpoT
    \nu_{f'}^{\dag}\nu_{f'}\left( {E',t} \right)
         } \right)
\quad ,
\label{5p2}
\eeq
where the tilde coefficients are defined by
\beq
  \tilde B_{\nu_f} \EoT \equiv
         B_{\nu_f} { { \lrb{aT}^3 } \over {2\pi^2}}
   {\left( {\fr{E}{T} } \right)}^2
\ , \quad \quad
  \tilde C_{\nu_f\nu_{f'}} \EoTEpoT \equiv
  {{E^2} \over {E'^2}} C_{\nu_f\nu_{f'}} \EoTEpoT
\quad .
\label{5p3}
\eeq
The $A$ coefficients remain unchanged.
Since $a^3 T^3$ is time independent,
so is $\tilde B_{\nu_f} \EoT$.

The second step is to convert to the vector formulation.
To accompany the $1$, $2$ and $3$ neutrino-vector components
specified in Sect.\ III,
we define new $0$ and $4$ components by
\beq
  v_0 \EoTt \equiv
  \left( {\nu_e^{\dag}\nu_e +
   \nu_\mu^{\dag}\nu_\mu } \right)\left( E, t \right)
\ , \quad \quad
  v_4 \EoTt \equiv
   \nu_\tau^{\dag}\nu_\tau \left( E, t \right)
\quad .
\label{5p4}
\eeq
and
\beq
  w_{\bar 0} \EoTt \equiv
   \left( {\bar \nu_e^{\dag}\bar \nu_e +
   \bar \nu_\mu^{\dag}\bar \nu_\mu } \right)
         \left( E, t \right)
\ , \quad \quad
  w_{\bar 4} \EoTt \equiv
   \bar \nu_\tau^{\dag}\bar \nu_\tau
         \left( E, t \right)
\quad .
\label{5p5}
\eeq
Here and henceforth, we use a bar
over a neutrino-vector index
to help distinguish neutrinos from antineutrinos.
Also,
to render the variable dependence the same
for production and oscillations,
we use the arguments $E/T$ and $t$ instead of $E$ and $t$
for the new components
$v_0$, $v_4$, $w_{\bar 0}$ and $w_{\bar 4}$.
The zero component $v_0$ is associated with
the electron-muon neutrino excess
in the canonical comoving volume,
while the fourth component $v_4$ is
the tau-neutrino excess in this volume.
Corresponding statements hold for the antineutrinos.

When reexpressed in vector notation,
Eq.\ \bref{2p8}
for pure production becomes
$$
  {{\prt v_a} \over {\prt t}} \EoTt  =
    {{4G_F^2T^5} \over {\pi^3}}
   \left( {
    -\sum\limits_b A_{ab} \EoT v_b \EoTt +
   B_a \EoT \delta \left( t \right)
          } \right.
\qquad\qquad
\qquad\qquad
$$
\beq
   + \sum\limits_b \int\limits_0^\infty
\left. {
           {{{ \dif E'} \over T}}
     C_{ab} \EoTEpoT
     v_b \EpoTt +
   \sum\limits_{\bar b} \int\limits_0^\infty
           {{{ \dif E'} \over T}}
 C_{a\bar b} \EoTEpoT
    w_{\bar b} \EpoTt
       } \right)
\quad ,
\label{5p6}
\eeq
where
$a = 0, 1,$ or $3$,
$b = 0, 1,$ or $3$
and
$\bar b = \bar 0, \bar 1,$ or $\bar 3$.
The new coefficients
$A_{ab}$, $B_a$, $ C_{ab}$,
and $ C_{a \bar b}$
in this equation are linear combinations of
the previously defined coefficients
$A_{\nu_f}$, $\tilde B_{\nu_f}$,
and $ \tilde C_{\nu_f \nu_f'}$.
They are given in Appendix B.
The new coefficients
are functions of neutrino energy and temperature,
but again only in the combinations $E/T$ and $E^\prime/T$.
They appear due to the change of basis
from the vectors
$\vec e_1$ and $\vec e_2$
to the vectors $\vec e_1 + \vec e_2$
and $\vec e_1 - \vec e_2$.

A similar derivation holds for antineutrinos,
yielding
$$
  {{\prt w_{\bar a}} \over {\prt t}} \EoTt =
   {{4G_F^2T^5} \over {\pi^3}}
   \left( {
    -\sum\limits_b A_{\bar a\bar b} \EoT w_{\bar b} \EoTt +
   B_{\bar a} \EoT \delta \left( t \right)
          } \right.
\qquad\qquad
\qquad\qquad
$$
\beq
  + \sum\limits_b \int\limits_0^\infty  {{{ \dif E'} \over T}}
        C_{\bar ab} \EoTEpoT
            v_b \EpoTt +
   \sum\limits_{\bar b} \int\limits_0^\infty
\left. {
         {{{ \dif E'} \over T}}
   C_{\bar a\bar b} \EoTEpoT
     w_{\bar b} \EpoTt
        } \right)
\quad .
\label{5p9}
\eeq
The coefficients $A_{\bar a\bar b}$ and $B_{\bar a}$
$C_{\bar ab}$ and $C_{\bar a \bar b}$
are also provided in Appendix B.
Note that if the first index of a coefficient is unbarred
then it enters in a neutrino equation.
If instead the first index is barred,
then the coefficient enters in an antineutrino equation.
Equations \bref{5p6} and \bref{5p9}
complete the change of basis to vector notation
for the pure-production case discussed in Sect.\ II.

The equations for the combined system
with simultaneous production and oscillations
follow from the above analysis.
For neutrinos,
adding the right-hand sides
of Eqs.\ \bref{3p2} and \bref{5p6} gives
$$
  {{\prt v_a} \over {\prt t}} \EoTt =
  \lrb{\vec v \EoTt \times \vec B_v }_a
\quad\quad
\quad\quad
\quad\quad
\quad\quad
\quad\quad
\quad\quad
\quad\quad
\quad\quad
$$
$$
 + {{4G_F^2T^5} \over {\pi^3}}
   \left( {
   -\sum\limits_b A_{ab} \EoT
     v_b \EoTt +
    B_a \EoT \delta \left( t \right)
         } \right.
\quad\quad
\quad\quad
$$
\beq
\left. {
   + \sum\limits_b \int\limits_0^\infty
         {{{ \dif E'} \over T}}
         C_{ab} \EoTEpoT v_b \EpoTt +
    \sum\limits_{\bar b} \int\limits_0^\infty
         {{{ \dif E'} \over T}}
    C_{a\bar b} \EoTEpoT w_{\bar b} \EpoTt
       } \right)
\quad .
\label{5p12}
\eeq
For antineutrinos,
adding the right-hand sides
of Eqs.\ \bref{3p2} and \bref{5p9} gives
$$
  {{\prt w_{\bar a}} \over {\prt t}} \EoTt =
   \left( {\vec w \EoTt \times
           \vec B_w} \right)_{\bar a}
\quad\quad
\quad\quad
\quad\quad
\quad\quad
\quad\quad
\quad\quad
\quad\quad
\quad\quad
$$
$$
   + {{4G_F^2T^5} \over {\pi^3}}
   \left( {
      -\sum\limits_b
   A_{\bar a\bar b} \EoT w_{\bar b} \EoTt +
     B_{\bar a} \EoT \delta \left( t \right)
          } \right.
\quad\quad
\quad\quad
$$
\beq
\left. {
   + \sum\limits_b \int\limits_0^\infty
         {{{ \dif E'} \over T}}
     C_{\bar ab} \EoTEpoT v_b \EpoTt  +
   \sum\limits_{\bar b} \int\limits_0^\infty
         {{{ \dif E'} \over T}}
    C_{\bar a\bar b} \EoTEpoT w_{\bar b} \EpoTt
       } \right)
\quad .
\label{5p13}
\eeq
Equations \bref{5p12} and \bref{5p13}
are the desired equations.
Note that in Eqs.\ \bref{5p12} and \bref{5p13}
we define
$$
  \left( {\vec v\left( E \right)
     \times \vec B_v } \right)_a \equiv
   0 \ , \quad \rm{ for } \ a=0 \ {\rm { and }} \ 4
\quad ,
$$
\beq
  \left( {\vec w\left( E \right)
    \times \vec B_w} \right)_{\bar a} \equiv
   0 \ , \quad  \rm{ for } \ \bar a=0 \ {\rm { and }} \ 4
\quad .
\label{5p14}
\eeq
The $1$, $2$ and $3$ components are computed normally.

\bigskip
\vglue 0.2cm
{\bf\large\noindent
VI.\ Technical Issues and Numerical Methodology}
\vglue 0.2cm
\setcounter{section}{6}
\setcounter{equation}{0}

In this section,
we first describe the discretization
of Eqs.\ \bref{5p12} and \bref{5p13}.
We next present some analytical results
for the resulting equations
and discuss the implementation of the
approximation for $\de(t)$,
already mentioned in Sect.\ II,
along with a method for partially compensating the
approximation of zero electron and positron mass.
Finally,
we present some details about the
algorithm used and the range of data obtained.

Numerical solution of
Eqs.\ \bref{5p12} and \bref{5p13}
requires conversion of the continuous energy variable
to a discrete one.
We allow $E/T$ to vary from
$\lrb{E/T}_{min}$ to $\lrb{E/T}_{max}$
and divide this interval into $N_E$ bins
of equal spacing.
Indices $j$ and $k$
ranging from $1$ to $N_E$
are used to label quantities associated
with neutrino and antineutrino energy bins,
respectively.
Most simulations were performed
with $\lrb{E/T}_{min} = 0.1$,
$\lrb{E/T}_{max} = 20.0$, and $N_E = 61$.

We define discrete density-like variables
associated with the $j$th and $k$th bins by
\beq
  \nu_f^{j \dag} \nu_{f'}^j =
  \nu_f^{\dag}   \nu_{f'} \left( {E^j} \right)
    {{\Delta E^j} \over T}
\quad ,
\quad\quad
  \bar \nu_f^{k \dag} \bar \nu_{f'}^k =
  \bar \nu_f^{\dag}   \bar \nu_{f'}
  \left( {E^k} \right){{\Delta E^k} \over T}
\quad .
\label{6p1}
\eeq
With this definition,
$  \nu_f^{j \dag} \nu_{f}^j $
represents the number of neutrinos
of flavor $f$
with energies between $E^j$ and $E^j + \Delta E^j$
in the canonical comoving volume $a^3 (t)$,
while
$ \bar \nu_f^{k \dag} \bar \nu_{f}^k $
is the number of antineutrinos of flavor $f$
in the $k$th energy bin in the volume $a^3 (t)$.
Discrete neutrino and antineutrino vectors
associated with the $j$th and $k$th bins
can similarly be introduced,
via
\beq
         v_a^j \lrb{ t } \equiv
   {{\Delta E^j} \over T} v_a        \EoTt
\ , \quad \quad
  w_{\bar a}^k \lrb{ t } \equiv
   {{\Delta E^k} \over T} w_{\bar a} \EoTt
\quad .
\label{6p2}
\eeq

Differential equations
for $v_a^j$ can be obtained by
multiplication of
Eq.\ \bref{5p12}
with ${{\Delta E^j}/T}$
and replacement of the continuous integral
over $E' / T$ by a discrete sum.
Since ${{\Delta E^j} /T}$
is time independent,
it can be moved inside the time derivative
on the left-hand side.
For antineutrinos
a similar procedure is performed
to give equations for $w_{\bar a}^k$.
The resulting discrete equations resemble
the continuum equations \bref{5p12} and \bref{5p13}
with the identifications
$$
  E \to E^i
\ , \quad \quad  E' \to E^m
\quad ,
\quad\quad
   \int\limits_0^\infty  {{{ \dif E'} \over T}} \to
    \sum\limits_m
\quad ,
$$
$$
   v_a        \lrb{ \frac{E^j}{T}, t }  \to v_a^j (t)
\ , \quad \quad
   w_{\bar a} \lrb{ \frac{E^k}{T}, t }  \to w_{\bar a}^k (t)
\quad ,
$$
$$
  A_{cd} \EoT \to
      A_{cd}^{i}
 \equiv A_{cd}\lrb{\frac{E^i}{T}}
\quad ,
\quad\quad
  B_c \EoT \to  B_c^i
 \equiv { {\Delta E^i} \over T} B_c\lrb{\frac{E^i}{T}}
\quad ,
$$
$$
  C_{cd} \EoTEpoT \to C_{cd}^{im}
 \equiv  {{\Delta E^i} \over {\Delta E^m}}
    C_{cd} \lrb{ \frac{E^i}{T} , \frac{E^m}{T} }
\quad ,
$$
$$
  \vec B_v \lrb{ {E , t} }
    \to \vec B_v^i (t) \equiv
  \vec B_v \lrb{ {E^i , t} }
\quad ,
\quad\quad
  \vec B_w \lrb{ {E , t} }
    \to \vec B_w^i (t)  \equiv
  \vec B_w \lrb{ {E^i , t} }
\quad ,
$$
$$
  V_{CP^+} \left( {E}   \right) \to V_{CP^+}^i \equiv
   V_{CP^+}\left( {E^i} \right)
\quad \quad \ {\mbox{ in }}
  \vec B_v {\mbox{ and }} \vec B_w
{\mbox{ of Eq.\ }\bref{3p3} }
\quad ,
$$
\beq
   \left\langle v_a (t) \right\rangle  \to
  \sum\limits_j v_a^j (t)
\ , \quad \quad
   \left\langle w_{\bar b} (t) \right\rangle  \to
  \sum\limits_k w_{\bar b}^k (t)
\quad \quad {\mbox{ in }} {{\vec V_{\nu \nu}} }
  {\mbox { of Eq.\ }\bref{3p6} }
\quad ,
\label{6p3}
\eeq
where the label
$c$ can be $a$ or $\bar a$ and
$d$ can be $b$ or $\bar b$,
and where the indices $i$ and $m$ stand for $j$ or $k$.

In the absence of production, \ie
setting $A_{cd}^{i} = B_c^i = C_{cd}^{im} = 0$,
the equations reduce to those
in ref.\ \ct{kps93a,ks94a}
for pure oscillations.
In the absence of oscillations, \ie
$\vec B_v = \vec B_w = 0$,
the discrete set of differential equations
can be analytically integrated,
as we show next.

The first step is to convert the time integration variable
$t$ to the neutrino temperature $T$,
via
\beq
  T = {1 \over {\left( {2\kappa_0 t} \right)^{1/2}}}
\quad ,
\quad\quad
  {{dT}\over{dt}} = -\kappa_0 T^3
\quad ,
\label{6p4}
\eeq
where
\beq
  \kappa_0 \equiv
    \left( {{{4\pi^3 G_N g_* } \over {45}}} \right)^{1/2}
\quad .
\label{6p5}
\eeq
Here, $G_N$ is Newton's constant and
$g_*$ counts the total number
of effectively massless degrees of freedom.
For $T > 1$ MeV,
$
  g_* = 10.75
$.

The structure of the discrete version of
Eqs.\ \bref{5p12} and \bref{5p13}
or of Eq.\ \bref{2p8} then becomes
\beq
  {{dx_r\left( T \right)} \over {dT}} =
   - k_0 T^2
    \left( {\calB_r\delta \left( T \right) +
     \calC_{rs} x_s\left( T \right)} \right)
\quad ,
\label{6p6}
\eeq
where $\calB$ is related to the $B$ coefficients
and $\calC$ contains both $C$ and $A$ coefficients.
Note that both the sets of coefficients
$\calB$ and $\calC$ are independent
of time and temperature.
In Eq.\ \bref{6p6},
\beq
  k_0 \equiv {{4G_F^2} \over {\kappa_0 \pi^3}}
\quad ,
\label{6p7}
\eeq
and $r$, $s$
represent both flavor and energy-binning indices:
$
  r \leftrightarrow c,i
$
with $c$ ranging over six values
and
$ i$ ranging from $1$ to $ N_E $.
In the formulation of Eqs.\ \bref{5p12} and \bref{5p13},
$x$ represents either $v$ or $\bar w$,
and $c$ ranges over
$
0,1,4, \bar 0,\bar 1,\bar 4
$.
For the formulation of Eq.\ \bref{2p8},
$x$ represents one of the
six distortions $\Delta_{\nu_f}$
and $c$ ranges over
$
  \nu_e,\nu_\mu ,\nu_\tau,
  \bar \nu_e,\bar \nu_\mu ,\bar \nu_\tau
$.
The solution to
Eq.\ \bref{6p6} is
\beq
  x_r \left( T \right) =
   k_0 \int_T^\infty  {d\bar T}~
   \bar T^2 \delta \left( {\bar T} \right)
   \left[ {
     \exp \left( { {{k_0} \over 3}
      \left( {\bar T^3-T^3} \right) \calC } \right)
          } \right]_{rs} \calB_s
\quad .
\label{6p8}
\eeq

Although the $6 N_E \times 6 N_E$ matrix $\calC$
is not symmetric,
it does have a complete set of right eigenvectors
$v^{\left( q \right)}$
defined by
\beq
   \calC v^{\left( q \right)} =
  \lambda^{\left( q \right)} v^{\left( q \right)}
\quad ,
\label{6p9}
\eeq
where
$
  q = 1, \ldots , 6 N_E
$.
The $\lambda^{\left( q \right)}$ are
the right eigenvalues of $\calC$.
The $v^{\left( q \right)}$
are not orthogonal but they are complete.
Thus, the vector $\calB$ can be expanded
in terms of the $v^{\left( q \right)}$ using
\beq
  \calB =
   \sum\limits_q
   \beta_{\left( q \right)}
        v^{\left( q \right)}
\quad ,
\label{6p10}
\eeq
where the $\beta_{\left( q \right)}$ are expansion coefficients.
The solution \bref{6p9} becomes
\beq
  x_r \left( T \right) =
   k_0\sum\limits_q
   \beta_{\left( q \right)}
   v_r^{\left( q \right)}
   \int_T^\infty  {d\bar T}\bar T^2
    \delta \left( {\bar T} \right)
    \exp \left( { {{k_0} \over 3}
    \left( {\bar T^3 - T^3} \right)
   \lambda^{\left( q \right)}} \right)
\quad .
\label{6p11}
\eeq

Equation \bref{6p11}
provides an efficient way to obtain results
for the pure-production case.
The integrals in \bref{6p11}
can be computed rapidly by various methods.
One needs
$\Re \lambda^{\left( q \right)} < 0$
for convergence.
For different values of $N_E$,
we have computed the right eigenvalues
of $\calC$ and verified convergence for sufficiently
large $N_E$.
This analysis aided our
determination of the minimum size of $N_E$
necessary for numerically accurate results.
As a side remark,
we note that a reasonable approximation
to pure production
can be obtained by saturating the sum
in Eq.\ \bref{6p11}
with terms corresponding to those eigenvalues
with the largest real parts.

To integrate the distortion equations,
a formula for $\delta (t)$ is needed.
We follow ref.\ \ct{dt92a}
and approximate $\delta (t)$ by $\delta_0 (t)$,
given by
\beq
  \delta_0 (t) =
  \lrb{
     \fr{12}{ 4 + z^3 K_1 (z) + 4 z^2 K_2 (z)}
      }^{ 1 / 3} - 1
\quad ,
\label{6p12}
\eeq
where $ z = m_e / T_\gamma$
and $K_1(z)$ and $K_2(z)$ are modified Bessel functions.
Recall that $\delta_0 (t)$ is obtained
under the assumption that neutrinos
are thermally decoupled for all times.

The direct use of
Eq.\ \bref{6p12} for arbitrarily early times
leads to an inconsistency.
It implies an early-time behavior
\beq
  \delta_0 (t) \approx
   \fr{1}{36} \lrb{ \fr{m_e}{T} }^2
\quad ,
\label{6p13}
\eeq
for $T \gg m_e$.
If this expression
were used in the pure-production equations,
one would find that the early-time behavior of
$\Delta_{\nu_f} \EoTt $ is
\beq
  \Delta_{\nu_f} \EoTt \approx
   c_{\nu_f} \EoT \lrb{ \fr{1}{T} }^2
\quad ,
\label{6p14}
\eeq
where $c_{\nu_f}\EoT$ is a time-independent constant.
Since the total excess density is obtained
by multiplying by ${E^2 dE}/{2\pi^2} $
and integrating over $E$,
it would follow that the excess density
of electron neutrinos over muon neutrinos
increases at earlier times.
If this result were correct,
it would pose a severe numerical difficulty.
When both production and oscillations are treated,
the neutrino-neutrino potential
would be important for $T \lsim 100 $ MeV.
This would necessitate starting the integration of
Eqs.\ \bref{5p12} and \bref{5p13}
at $T \sim 100 $ MeV.
To integrate over the large temperature range
from $\sim 100$ MeV down to
$\sim 1$ MeV would be prohibitive in computer time.

Fortunately,
the early-time behavior
in Eq.\ \rf{6p13} is incorrect.
The source of the difficulty is the assumption
that neutrinos are thermally decoupled for all times.
It must be true that for $T > 10$ MeV
neutrinos are in thermal equilibrium with photons,
so that $\delta$ is exponentially small.
The results of refs.\ \ct{dt92a} and \ct{df}
suggest that neutrinos decouple in the temperature range
from $2$ MeV to $5$ MeV.
Higher-energy neutrinos decouple later,
at a lower temperature.
Also, electron neutrinos decouple somewhat later
than muon and tau neutrinos.
In any case,
it is evident that $\delta$ should be set to zero
at sufficiently early times.

An exact formula for $\delta (t)$
is not available.
Instead,
we make the simple approximation
$$
  \delta (t) = \delta_0 (t)
\ , \quad  \mbox{ for } t > t_s
\quad ,
$$
\beq
  \delta (t) = 0
\ , \quad  \mbox{ for } t < t_s
\quad ,
\label{6p15}
\eeq
where $t_s$ is a cutoff time.
Numerically, $t_s$ is the time
at which we start the integration
of Eqs.\ \bref{5p12} and \bref{5p13}.

In principle,
a better approximation can be made
by taking $t_s$ to be a function of bin energy.
In other words,
$\delta$ is set to zero earlier
for low-energy neutrinos
and later for high-energy neutrinos.
If one is interested only in results
for $t \gsim 0.3$ seconds,
then the detailed manner in which $\delta$
is turned on is unimportant.
For simplicity,
we take $t_s$ to be energy-bin independent.
We denote by $T_s$ the neutrino temperature
at the time $t_s$.

The approximation \bref{6p15} is a significant improvement
over the use of $\delta_0$ at all times.
It is also better than
incorporating the $\delta T_\gamma / T$ correction
to $\delta$ provided in ref.\ \ct{dt92a}.
For numerical purposes,
we take $T_s = 3.0$ MeV,
corresponding to a starting time $t_s$
of about $0.082$ seconds.
We have varied $T_s$ somewhat and checked
that final production results
are relatively insensitive to this choice.
Note that,
although it is important to turn off $\delta$
for early times in dealing with neutrino oscillations,
it is less important in the pure-production case
if one is interested in results at late times.
Hence, using the approximation $\delta_0$ at all times
is unlikely to affect the conclusions
of ref.\ \ct{dt92a}
concerning the primordial $^4$He abundance.

Another approximation made is to set
the electron and positron mass to zero.
When the temperature drops below $0.5$ MeV,
electrons and positrons quickly annihilate
and production rapidly ceases.
This is not correctly incorporated
in the formulae for the scattering processes
when $m_e = 0$ is used,
so some error is made below $1$ MeV.
A simple way to compensate for this
is to stop the production prematurely.
We chose to terminate our numerical integrations
involving production processes
at the temperature $T_f = 0.75$ MeV,
which corresponds to a time $t_f$
of about $1.312$ seconds.
Beyond $T_f$,
we proceed with integration
of the pure oscillation equations
\bref{3p2} instead.
Varying $T_f$ from $0.9$ to $0.5$ MeV
changes the distortion functions by about $20 \%$.
Hence,
the $20 \%$ uncertainty in selecting $T_f$
is the biggest uncertainty in our results.

To minimize the chances of programming errors,
at all stages two independent programs were written
and results were checked to double-precision machine accuracy.
Desktop Hewlett-Packard and Sun work stations were used.

The numerical integrations
were performed with a fourth-order Runge-Kutta algorithm
applied to Eqs.\ \bref{5p12} and \bref{5p13}.
The scale factor $a(t)$ at $T_s = 3.0$ MeV
was taken to be $0.5$ MeV$^{-1}$ to facilitate comparison
with the results of ref.\ \ct{ks94a}.
The point is that when $T_s = 1.5$ MeV,
$a (t)$ becomes $1.0$ MeV$^{-1}$,
which is the value selected in ref.\ \ct{ks94a}.
Since the $A$, $B$, and $C$ coefficients are time-independent,
they could be computed once initially and stored in memory.

The integration combining production and oscillation
proceeds about one-hundred times slower than
pure oscillation integration
due to the increased amount of arithmetic.
For this reason,
verification of all the runs
in ref.\ \ct{ks94a} was impractical.
However,
sufficiently many simulations
were performed to confirm the conclusions
of ref.\ \ct{ks94a}.
In addition to the situation without oscillations,
we studied the cases
$\ss = 0.81$ with
$\Delta = 10^{-12}$ eV$^2$,
$\Delta = 10^{-9}$ eV$^2$,
$\Delta = 10^{-7}$ eV$^2$,
$\Delta = 10^{-6}$ eV$^2$,
and $\Delta = 10^{-4}$ eV$^2$,
and the cases
$\ss = 0.25$ with
$\Delta = 10^{-12}$ eV$^2$,
$\Delta = 10^{-9}$ eV$^2$,
and $\Delta = 10^{-6}$ eV$^2$.

\bigskip
\vglue 0.2cm
{\bf\large\noindent
VII.\ Results for Pure Production without Oscillations}
\vglue 0.2cm
\setcounter{section}{7}
\setcounter{equation}{0}

For the case without flavor mixing,
the improvements considered in Sect.\ VI
lead to somewhat different
nonequilibrium distortion profiles
than those of
refs.\ \ct{dt92a,df}.
This section presents our results and
discusses these differences.

Figure 1
displays our final production profiles at $T_f = 0.75$ MeV
for zero mixing angle.
The curves plotted are
$\nu_e^{\dag} \nu_e (E) / T $,
$\nu_\mu^{\dag} \nu_\mu (E) / T $,
their sum, and their difference,
as functions of neutrino energy $E$.
The continuous distortion densities
$\nu_f^{\dag} \nu_f (E) $ are defined
in Eq.\ \bref{5p1}.
We remind the reader
that $\nu_f^{\dag} \nu_f (E)  \Delta E / T$
represents the excess number of neutrinos of flavor $f$
between $E$ and $E + \Delta E$
in the canonical comoving volume.
Each curve therefore represents
a neutrino excess per unit energy
over the Maxwell-Boltzmann distribution
in the volume $a^3$,
and the area under a curve is the total excess
in that volume.

The production profile for the tau neutrinos
is identical to that for the muon neutrino
because,
in the absence of mixing,
the system is symmetric under the interchange
of muon and tau flavors.
Also,
the distortions for antineutrinos
are the same as for the corresponding neutrinos
because
the system is CP symmetric when $\th = 0$.

The electron- and muon-neutrino curves are
negative for small $E$,
indicating a deficit of neutrinos in the low-energy region
compared to the Maxwell-Boltzmann distribution.
Physically,
the effect arises from
the reheating of neutrinos by $e^+$-$e^-$ annihilation,
which shifts some low-energy neutrinos to higher energies
and thereby reduces the number in the low-energy region.
As can be seen from Figure 1,
the deficit largely cancels
in the neutrino-difference profile,
which represents the difference between
electron- and muon-neutrino number densities.
The flavor-density nonequilibrium distortions
peak at $E \approx 4 T$,
while the difference curve peaks at $E \approx 3.3 T$.
All curves have exponential tails
that become small for $E \gsim 13 T$.

The distortion curves in Figure 1
evolve with time for $t > t_f$
due to the expansion of the Universe
and the associated energy redshifts.
For $T < 0.75$ MeV,
the curves maintain the same overall shape
but are compressed horizontally
and expanded vertically
by the factor $T_f / T$,
where $T_f = 0.75$ MeV.
Thus, at later times the peaks at $\sim 3.0$ MeV
approach the origin and increase in height.
The areas under the curves remain constant,
since neutrinos are neither created nor destroyed
when $t > t_f$.

It is useful to have analytical
expressions representing the pure-production profiles
in Figure 1.
We find the electron- and muon-neutrino profile
curves are accurately reproduced by
the expressions
$$
  \nu_e^{\dag} \nu_e (E) / T
 \approx
  \left( { -1.3567 \times 10^{-2} + 1.9692 \times 10^{-2} x
       + 2.4600 \times 10^{-3} x^2 }\right.
$$
$$
  \left. {  + 5.9818 \times 10^{-5} x^3
    - 7.5683 \times 10^{-6} x^4 } \right)
 \fr{ x^2 }{2 \pi^2} \exp (-x)
\quad ,
$$
$$
  \nu_\mu^{\dag} \nu_\mu (E) / T
 \approx
  \left( { -1.0568 \times 10^{-2} + 1.0668 \times 10^{-2} x
       + 1.4263 \times 10^{-3} x^2 }\right.
$$
\beq
  \left. {  + 9.1969 \times 10^{-5} x^3
    - 7.5711 \times 10^{-6} x^4 } \right)
 \fr{ x^2 }{2 \pi^2} \exp (-x)
\quad ,
\label{7p2}
\eeq
where $x = E/T$.

Figure 2 presents a comparison
of our electron-neutrino and neutrino-difference profiles
with the estimate of ref.\ \ct{df},
which is
\beq
\nu_e^{\dag} \nu_e (E) / T =
 \lrb{6 \times 10^{-4}} \fr{E}{T}
  \lrb{\fr{11 E}{ 4 T} - 3}
  \fr{E^2}{2 \pi^2} e^{-E/T} a^3
\quad .
\label{7p1}
\eeq
Our results exhibit two principle differences
from those of ref.\ \ct{df}.
First,
the analytical approximation \rf{7p1}
somewhat overestimates the electron-neutrino excess.
In the canonical volume $a^3$
with $a (t) = 2.0$ MeV$^{-1}$ at $t = t_f$,
we obtain a total electron-neutrino excess
of about $2 \times 10^{-3}$.
In contrast,
the analytical approximation \rf{7p1}
gives about $4.8 \times 10^{-3}$.
Second,
our distortions reach a maximum
at a sightly lower energy.
The peak in our neutrino-difference profile
occurs at an energy value about $25 \%$ below
that of the peak in Eq.\ \bref{7p1}.

The total muon-neutrino excess
we obtain in the canonical volume is
about $1 \times 10^{-3}$
so that the total excess of electron neutrinos
over muon neutrinos
is also about $1 \times 10^{-3}$.
Since the oscillation simulations in
refs.\ \ct{kps93a,ks93a,ks94a}
were based on the initial production profile \bref{7p1},
our current simulations lead to some modifications
of the earlier results.
These effects are discussed in Sect IX.

Figure 3 shows the time development of
the nonequilibrium distortions.
Various combinations of components
of the average neutrino vector
are plotted as functions of time.
The curves begin at $t_s \approx 0.082$ seconds
and end at the final production time
of $t_f \approx 1.312$ seconds.
The production proceeds smoothly and monotonically.
The total number of excess
electron neutrinos in the canonical volume
corresponds to $\vev{\lrb{v_0 + v_1}/2}$,
while the total number of excess
muon neutrinos
orresponds to $\vev{\lrb{v_0 - v_1}/2}$.
The difference between electron and muon neutrinos
is given by the component $\vev{v_1}$,
while their sum is $\vev{v_0}$.
The total excess of tau neutrinos is
the same as for muon neutrinos,
and hence is given by $\vev{v_4}=\vev{\lrb{v_0 - v_1}/2}$.
Again, CP symmetry
ensures that the results
for antineutrinos are the same as those for neutrinos.

Overall,
our results for pure production are similar to those of
ref.\ \ct{dt92a}.
The final neutrino distortion productions in ref.\ \ct{dt92a}
were terminated at $0.1$ MeV.
When extrapolated to $T_f$,
they are about $35 \%$ larger
than ours but are otherwise comparable.
As previously noted,
the difference lies in the treatment
of the electron mass.
However,
since we do not assume neutrinos
are in thermal equilibrium at very early times,
our early-time results differ.
For instance,
taking into account thermalizing neutrino interactions,
we find that the $T = 8$ MeV electron-neutrino distortion
$\Delta_{\nu_e}$ is virtually zero,
unlike the results in Figure 4 of ref.\ \ct{dt92a}.
For the same reason,
we find a much smaller profile at $T = 4$ MeV.
Similarly,
our results for Figures 5--8 of ref.\ \ct{dt92a}
produce curves dropping sharply for $T \gsim 4$ MeV,
without the $1/T^2$ behavior for early times.

\bigskip
\vglue 0.2cm
{\bf\large\noindent
VIII.\ Results for Production with Oscillations}
\vglue 0.2cm
\setcounter{section}{8}
\setcounter{equation}{0}

This section discusses our numerical results
for the production of nonequilibrium distributions
when neutrino mixing is present.

For $\Delta < 10^{-8}$ eV$^2$,
our results resemble the $\theta = 0$ case
discussed in Sect.\ VII.
This follows because the vacuum term is small
for $t_s < t < t_f $ and $\Delta < 10^{-8}$ eV$^2$.
The neutrino vectors, antineutrino vectors,
and effective magnetic fields
are all directed near the flavor axis, \ie
the $1$-axis.
Since they point in a common direction,
the cross-product terms
$\vec v \times \vec B_v$
and $\vec w \times \vec B_{w}$
in Eqs.\ \bref{5p12} and \bref{5p13}
are close to zero,
and production proceeds as if
there were no neutrino mixing.

When $\Delta > 10^{-8}$ eV$^2$,
some excess electron neutrinos
convert to muon neutrinos during the production phase.
As expected,
the effect is greater for larger $\theta$.
As a consequence of the neutrino conversion,
the electron-neutrino and difference profiles are reduced
while the muon-neutrino profile is increased.

Figure 4 displays the results
for $\Delta = 10^{-6}$ eV$^2$
and $\sin^2 2 \theta = 0.81$.
The curves representing the electron-
and muon-neutrino sum profile and the
tau-neutrino profile are almost identical to
those in Figure 1.
This also holds true at other values of $\De$ and $\th$.
The reason is that the oscillation terms
do not directly enter the expressions for these profiles,
as can be seen from
Eqs.\ \bref{5p12} and \bref{5p13}.
Thus,
oscillations affect them only by feedback
from the conversion among electron and muon neutrinos
into the scattering processes in the Boltzmann equation.
This secondary effect is relatively small.

We have constructed an analytical approximation
to the production profiles when $0 < \theta < \pi /4$,
given the pure-production results.
It is based on the following idea.
In a small time interval during production,
a new excess of electron neutrinos is supplied
to the system.
This corresponds to adding small components
to neutrino vectors along the $1$-axis.
These small additional components
rotate around the associated magnetic fields.
Since the production is continuous
and since the time scale for oscillations is
much shorter than the time scale for production,
when averaged over an oscillation time
the new small components lead to contributions
aligned with the magnetic fields.
Thus, ANME configurations are achieved.
The averaging effect is equivalent
to projecting the production vectors
onto their magnetic fields.
To construct analytical ANME configurations,
we exploit the scheme presented in ref.\ \ct{ks94a}
and mentioned in Sect.\ IV:
the partial cancellation
between the neutrino-neutrino and CP-asymmetric terms
means that reasonably accurate ANME configurations
can be obtained using only the vacuum and CP-symmetric terms.
Then,
for each $E$ one projects the new production contributions
onto the analytical ANME configurations.
This must be done continuously
during the production phase,
and the results summed.

It is useful to find a simpler approach based
on the final pure-production profile.
Then,
one can employ the fits in
Eq.\ \bref{7p1}.
In much of the parameter region of our simulations,
the magnetic fields governing neutrino oscillations
are changing.
Initially,
the magnetic fields point along the $1$-axis.
Hence, ANME configurations are flavor eigenstates.
For $\Delta > 10^{-8}$ V$^2$,
magnetic fields rotate toward $\vec \Delta$
during the later stages of production.
If most of the production
occurs before this rotation,
then the production profile for $\th \ne 0$
is obtained by \it rotating \rm
the pure-production case,
for which $\th =0$, onto ANME configurations.
If most of the production occurs after the rotation,
then the production profile for $\th \ne 0$
is obtained by \it projecting \rm
the pure-production case onto ANME configurations.
In general,
early production is rotated onto ANME configurations,
while later production is projected.
Hence,
we expect the true production profile
to lie between the projected and rotated curves.

The analytical approximation to the data
is implemented as follows.
It is understood that all vectors
in this paragraph have three components.
Define the CP-symmetric part $\vec B_+$
of the magnetic field by
\beq
\vec B_+ \lrb{E, t} =
   {{\vec \Delta } \over {2E}} -
   {V_{CP^+} (E)}  \hat e_1
 \quad .
\label{8p1}
\eeq
Let $\vec v^{(0)} \lrb{E, t_f}$ be the vector obtained
in numerical simulations
for pure production with $\th = 0$.
Then,
the rotated vector
$\vec v_r \lrb{E, t_f}$
is given by
\beq
 \vec v_r \lrb{E, t_f} =
  \pm \fr{\vert \vec v^{(0)} \lrb{E, t_f} \vert }
         {\vert \vec B_+     \lrb{E, t_f} \vert }
   \vec B_+ \lrb{E, t_f}
\quad ,
\label{8p2}
\eeq
where the $\pm$ sign is the sign of
$\vec v^{(0)} \lrb{E, t_f} \cdot \vec B_+ \lrb{E, t_f}$.
The projected vector
$\vec v_p \lrb{E, t_f}$
is given by
\beq
 \vec v_p \lrb{E, t_f} =
  \fr{\vec v^{(0)}   \cdot \vec B_+ \lrb{E, t_f} }
     {\vec B_+ \cdot \vec B_+ \lrb{E, t_f} }
   \vec B_+ \lrb{E, t_f}
\quad .
\label{8p3}
\eeq

The analytical approximation based on rotation,
with data approximated using Eq.\ \bref{8p2},
works quite well for small $\Delta$ or small $\th$.
For example,
when $\Delta < 10^{-7}$ eV$^2$,
the agreement for
$v_1 \lrb{E, t_f}$ and $v_2 \lrb{E, t_f}$
is at the one-percent level
even for the large-angle case with $\ss= 0.81$.
The same accuracy is obtained
for the case with
$\Delta = 10^{-6}$ eV$^2$ with $\ss=0.25$.

Figures 5a and 5b
show profile plots of the components $v_1$ and $v_2$
for the case $\Delta = 10^{-6}$ eV$^2$, $\ss=0.81$.
As expected,
the data lie between the two analytical approximations
based on rotation and on projection.
In the region where $E$ is a few MeV,
the analytical rotation method overestimates
the vectors by about $10 \%$.
The error reduces to about $5 \%$
in the energy range near $8$ MeV,
while for $E \gsim 11$ MeV
the method is accurate to at least one percent.
The higher-energy vectors are reproduced better
because their magnetic fields
undergo relatively less rotation.

The analytical approximation also works
at earlier times during the production process.
The idea is to use
Eq.\ \bref{8p2} or \bref{8p3}
with the final time $t_f$
replaced by an arbitrary time $t$
in the production interval $t_s < t \le t_f$.
For example,
consider the time $t \approx 0.5$ seconds
in the middle region of the production phase
for the case with
$\Delta = 10^{-4}$ eV$^2$ and $\ss=0.81$.
As expected,
our data show that the values of
$v_1$ and $v_2$ again fall between
the curves obtained from
the rotation and projection approximations.
For $E > 4$ MeV,
the rotation method fits
the data to about $10 \%$.
For $E < 4$ MeV,
the rotation results are larger than the data,
with significant deviations when $E < 2$ MeV.
For $E < 2$ MeV,
the projection method underestimates
the data by about $30 \%$.
The projection approximation is better
for low-energy neutrinos here
because the corresponding ANME configurations
undergo significant early rotation during production.

In the parameter region
with $\Delta > 10^{-4}$ eV$^2$,
the full simulation of the production phase
involves prohibitive computer time.
The production profile should
lie closer to the projection method
based on Eq.\ \bref{8p3}.
For $\Delta > 10^{-2}$ eV$^2$,
the projection approximation should be reasonably accurate.
Indeed,
in a test run for
$\ss = 0.81$ and $\Delta = 10^{-1}$ eV$^2$,
the data were reproduced within statistical uncertainties.

The analytical approximation
and the insights obtained in our earlier work,
summarized in Sect.\ IV,
suggest several predictions for
neutrino behavior during the production phase.
First, neutrinos should maintain themselves
in ANME configurations.
Second, the flavor development of neutrinos
should be smooth.
This follows from the ANME property
and the slow evolution of the effective magnetic fields
during the production time interval from $t_s$ to $t_f$.
Third, alignment should hold.
This is a consequence of the ANME property
and the approximate alignment of the
effective magnetic fields.
Fourth, there should be planarity.
This follows because
instantaneous nonlinear mass eigenstates have $ v_3 = 0$
and so,
since the neutrinos are in ANME configurations,
$v_3$ should be much smaller than $v_1$ and $v_2$.
Finally, CP asymmetry should be suppressed.
The ANME property implies the existence of
the mechanism for CP-asymmetry cancellation
given in ref.\ \ct{ks94a}.

Data obtained during our production runs
confirm these five predictions.
All the data display
smooth neutrino-flavor behavior.
Figure 6 shows the components of the average
neutrino vector as a function of time for the case with
$\Delta = 10^{-6}$ eV$^2$ and $\ss=0.81$.
The smooth behavior is evident.
The planarity feature is also evident
since the third component is small.
In fact,
the third component is at least three orders
of magnitude smaller than the first component.
The antineutrino-vector components
are omitted from Figure 6
because they are indistinguishable
from the vector components on the scale of the figure.
The difference between neutrino and antineutrino vectors
is typically four or more orders of magnitude
smaller than the vectors themselves.
This reflects the CP-suppression mechanism.
We also examined data taken during the
$\Delta = 10^{-4}$ eV$^2$ and $\ss=0.81$ production cycle,
which is expected to exhibit the most sensitive
dependence on the ANME and alignment properties.
The data reveal that these properties hold well.
The test results are similar to those shown in Table IV
of ref.\ \ct{ks94a}.

\bigskip
{\bf\large\noindent
IX.\ Results for Pure Oscillations after Production}
\vglue 0.2cm
\setcounter{section}{9}
\setcounter{equation}{0}

This section describes results
for oscillations governed
by Eq.\ \bref{3p2}
but
based on the final production profile
obtained from the Boltzmann equation
with oscillations.
We also discuss some differences
with results in our earlier works
\ct{kps93a,ks93a,ks94a},
in which the production profile
is treated in a step-function approximation
based on the analytical approximation
of ref.\ \ct{df}
given by
Eq.\ \bref{7p1}.
For purposes of comparison
with the earlier work,
we introduce in this section
the normalized vectors $\vec r_v (t)$
defined by
\beq
 \vec r_v (t) =
 \fr{ \vev {\vec v (t)}}{\vev{\vert\vec v (t) \vert }}
\quad .
\label{9p1}
\eeq
In this equation,
$ \vev{\vert\vec v (t)   \vert} =
 \sum_j\vert\vec v^j (t) \vert$
is the total number of neutrinos
in the canonical comoving volume at time $t$,
which is constant during the oscillation phase
$t > t_f$ because neutrinos are
neither created nor destroyed in the comoving volume.

Figure 7
displays the components $r_{v1}$ and $r_{v2}$
as a function of time for the case with
$\Delta = 10^{-12}$ eV$^2$ and $\ss = 0.81$.
The solid lines represent our new data,
while the dashed lines represent data
from earlier work \ct{ks94a}.
The figure shows that a flavor conversion
occurs about $5$ seconds earlier in our new data.

The qualitative features of the two sets of data
agree.
In both sets of data,
the behavior is evolutionary and smooth.
At early times with $t < 40$ seconds,
neutrinos are in approximate flavor eigenstates
and hence the neutrino vectors
lie along the $1$-axis.
The relatively small initial vacuum term grows with time
and eventually dominates,
so that for $t > 120$ seconds
the neutrinos are instead
in approximate vacuum-mass eigenstates
with their vectors lying along $\vec \Delta$.
Both sets of data display a feature called
precocious rotation,
\ie the rotation
to vacuum-mass eigenstates occurs significantly
earlier than in the absence of the nonlinear term.
The results differ significantly
from data taken in the absence of neutrino-neutrino interactions.
Without the nonlinear term,
magnetic fields are dominated by the vacuum term
only for $t \gsim 500$ seconds,
damped oscillations occur,
and neutrinos attain vacuum behavior
only after $t > 1500$ seconds.

The reason for the earlier flavor conversion
in the new data is the use of
our new production profile.
As noted in the discussion below
Eq.\ \bref{7p1},
there are two main differences
between our new profile and the one
obtained in ref.\ \ct{df}:
the total production is smaller,
and the average energy is lower.
See Figure 2.
The change in the total production
largely cancels in the ratio $\vec r_v$.
It is the energy shift that is responsible
for the $5$-second time difference
appearing in Figure 7.
The partial cancellation between the terms
$\vec V_{\nu \nu} $ and ${V_{CP^-}} $
means that the precocious rotation occurs when
$\vert { {{\vec \Delta } /{2E}} } \vert $ and
$\vert {  V_{CP^+} (E) } \vert $
become approximately equal
\ct{ks94a}.
When the average energy of the profile is smaller,
the vacuum term is larger and so the rotation
to vacuum-mass eigenstates occurs sooner.

A similar effect appears for other values
of $\De$ and $\ss$.
For example,
the curves for $r_{v1}$ and $r_{v2}$
in the case with
$\Delta = 10^{-9}$ eV$^2$ and $\ss = 0.81$
are shifted relative to our earlier results
by about one second in the crossover region
between $3$ and $16$ seconds.
This is again due
to the smaller average energy of the profile,
which causes an early rotation to vacuum-mass eigenstates
because ANME configurations align
sooner with $\vec \Delta$.
Beyond $16$ seconds, the two sets of data coincide.
Likewise,
the curves for $r_{v1}$ and $r_{v2}$
in the case with
$\Delta = 10^{-6}$ eV$^2$ and $\ss = 0.81$
display time shifts of about $0.1$ seconds
in the crossover region between $0.3$ and $1.5$ seconds.
Beyond $2.0$ seconds, results again coincide.
For cases with smaller mixing angles,
the time-difference effect is comparable
but is less pronounced in plots because
the neutrino flavor changes less.
We remark in passing that the
effect is absent for the case with
$\Delta = 10^{-4}$ eV$^2$ and $\ss = 0.81$
because the overlap between the new data
and those of ref.\ \ct{ks94a} occurs
when neutrinos are already near vacuum-mass eigenstates.

The results we have obtained confirm all the qualitative
conclusions of ref.\ \ct{ks94a},
summarized in Sect.\ IV.
The numerical results of \ct{ks94a} are accurate,
except for the above-mentioned time-shift effect
in the crossover region.

The analytical approximations obtained in
the present work and in ref.\ \ct{ks94a}
provide a means to determine neutrino vectors
to an accuracy better than the intrinsic
methodological uncertainties discussed in Sect.\ II.
There are two stages involved.
First,
given a value of $\Delta$ and $\ss$,
the neutrino-distortion production profile
can be determined via the techniques
provided in Sect.\ VIII.
Second,
results for post-production times $t > t_f$
can be found using
the method of Sect.\ VI.E in ref.\ \ct{ks94a}.
This method gives the neutrino vectors as
\beq
 \vec v \lrb{E, t } \approx
   \pm \fr{\vert { \vec v   \lrb{E, t_f} } \vert}{
           \vert { \vec B_+ \lrb{E, t  } } \vert }
    \vec B_+ \lrb{E, t}
\quad ,
\label{9p2}
\eeq
where $\vec B_+ \lrb{E, t}$
is defined in Eq.\ \bref{8p1},
and where the overall sign is
that of the dot product
$\vec v \lrb{E, t_f } \cdot \vec B_+ \lrb{E, t_f }$.
The antineutrino vectors
are set equal to the neutrino vectors,
$ \vec w \lrb{E, t} = \vec v \lrb{E, t}$,
thus guaranteeing perfect CP-suppression.
The third components of vectors are zero,
thus ensuring planarity.
The second stage of the method reproduces
numerical results to about two decimal places
\ct{ks94a}.
Via this two-step procedure,
reasonably accurate numbers
for all quantities can be obtained.

\bigskip
\vglue 0.2cm
{\bf\large\noindent
X.\ Summary}
\vglue 0.2cm
\setcounter{section}{10}
\setcounter{equation}{0}

In this work,
we have treated
the nonequilibrium thermodynamic statistics
of neutrinos in the early Universe
in the presence of neutrino mixing and oscillations.
Differential equations were derived
that determine the deviations
from standard equilibrium statistics.
They are Eqs.\ \bref{5p12} and \bref{5p13}.
We have resorted to numerical methods
to solve these equations,
since they are complicated and nonlinear.
We summarize here
the behavior of the solutions
and our methods for approximating them analytically.

Soon after the start of the production
of nonequilibrium distortions,
neutrinos and antineutrinos achieve
configurations that are
approximate nonlinear mass eigenstates (ANME).
As the Universe expands,
various interaction terms change.
However,
the neutrinos remain in ANME configurations
during these changes,
so the flavor content evolves smoothly.

For $\Delta < 10^{-8}$ eV$^2$,
the production of nonequilibrium distortions
occurs when the vacuum term is relatively small.
This implies that the production phase proceeds
as in the case without neutrino mixing,
producing neutrinos close to flavor eignenstates.
During the oscillation phase,
the vacuum term increases
while the interaction effects decrease.
Eventually, the vacuum term dominates.
Neutrinos therefore gradually evolve
from flavor eigenstates to vacuum-mass eigenstates.
For $\Delta < 10^{-9}$ eV$^2$,
the rotation occurs earlier than one might expect
if neutrino-neutrino interactions are neglected.
This precocious rotation,
observed and explained in ref.\ \ct{ks94a},
is confirmed in our simulations.

For $\Delta > 10^{-8}$ eV$^2$
oscillations play a significant role
already during the production phase.
Since the nonequilibrium distortions
are induced by the weak interactions,
the excess neutrinos are generated as flavor eigenstates.
However,
the time scale for oscillations
is much shorter than that for production,
so the neutrinos quickly oscillate, decohere,
and achieve ANME configurations.
For $\Delta < 10^{-7}$ eV$^2$
much of the production occurs
while ANME configurations point in the flavor direction.
Hence,
a good approximation to the final production profile
in this case
is to \it rotate \rm the final pure-production
(zero mixing) results onto ANME configurations.
For $\Delta > 10^{-2}$ eV$^2$
most of the production occurs
while ANME configurations point
in the mass-eigenstate direction.
Hence,
a good approximation to the final production profile
in this case is to \it project \rm the pure-production results
onto ANME configurations.
In the region
$ 10^{-7}$ eV$^2 < \Delta < 10^{-2}$ eV$^2$
these two approximations straddle
the true production results.

The above approximations to the final production profile
can be made analytical
by using the cancellation mechanism of ref.\ \ct{ks94a}
so that mass eigenstates are constructed
based solely on the vacuum term and
the $CP$-conserving interaction
between the charged leptons and the neutrinos.
For the parameter region of our simulations,
we find that the rotation method reproduces
the data extremely well
for $\Delta < 10^{-7}$ eV$^2$.
For $\Delta \approx 10^{-6}$ eV$^2$ and $\ss =0.81$,
it overestimates results by at most $10 \%$.
One can compensate for this effect.
We have also described in Section VIII
how to extend the analytical method to
the production-oscillation results
for larger $\Delta$.

During the oscillation phase
for $\Delta > 10^{-8}$ eV$^2$,
neutrinos continue to rotate
toward vacuum-mass eigenstates.
The behavior therefore continues to
be smooth and evolutionary.
For $\Delta > 10^{-2}$ eV$^2$,
we expect the production phase to
render neutrinos close to vacuum-mass eigenstates.
Since the vacuum term dominates thereafter,
neutrinos remain in vacuum-mass eigenstates
during the oscillation phase.
The result should be constant flavor behavior.
Indeed, short test simulations confirm this.

The improved treatment of the production phase
means that our results for \it unnormalized \rm quantities
are now accurate to within systematic uncertainties.
This represents an improvement in absolute accuracy
of a factor of about twenty.

The approximate analytical method
of ref.\ \ct{ks94a}
can be used to obtain to good precision
the results for the oscillation phase.
By combining this method
with the new analytical treatment of the production,
we obtain a complete analytical approximation
for the neutrino and antineutrino vectors
over the entire time period of interest.

All the qualitative neutrino flavor behavior
of refs.\ \ct{kps93a,ks94a}
has been confirmed.
Features including
smooth evolutionary behavior,
alignment, planarity,
maintenance of ANME configurations,
and CP suppression
are observed not only in the oscillation phase
but also during the production phase.

We have uncovered one quantitative difference
with ref.\ \ct{ks94a}:
neutrinos rotate to vacuum-mass eigenstates
about $5 \%$ earlier in time.
The reason is that
the approximate initial distortion distributions used
in ref.\ \ct{ks94a}
are shifted somewhat to higher energies
than those produced in our present numerical data.
This caused the vacuum term to dominate somewhat later
in the simulations of
ref.\ \ct{ks94a}.
Asymptotically,
the results of our current and previous work agree.

Within the parameter region of this paper,
we have found little generation of CP asymmetry.
This means that nonequilibrium distortions
are unlikely to affect nucleosynthesis substantially.
Although one might \it a priori \rm
expect neutrino oscillations
to have observable effects,
the CP-suppression mechanism we uncovered in our simulations
implies that the impact on the primordial helium abundance
is likely to be at the same level as found
in ref.\ \ct{dt92a},
where neutrino mixing was absent.

\bigskip
\vglue 0.2cm
\noindent
\secttit{Acknowledgments}

We thank Scott Dodelson and Mike Turner
for discussions and for making available to us
their pure-production data.
This work is supported in part
by the United States Department of Energy
(grant numbers DE-FG02-91ER40661 and DE-FG02-92ER40698),
by the Alexander von Humboldt Foundation,
and by the PSC Board of Higher Education at CUNY.

\bigskip
\vglue 0.2cm
{\bf\large\noindent
Appendix A: Coefficients
for Three-Flavor Pure Production}
\vglue 0.2cm
\setcounter{section}{11}
\setcounter{equation}{0}
\def\theequation{A.\arabic{equation}}

This appendix provides the $A$, $B$,
and $C$ coefficients appearing
in Eq.\ \bref{2p8}.
They are functions of the scaled initial
and final neutrino energies
$\frac{E}{T}$ and $\frac{E'}{T}$.
This functional dependence is dropped here
to save space.
In what follows,
it is useful to define the electroweak coupling constants
\beq
  a \equiv ( 1 + \sinsthw )^2
\ , \quad \quad
  b \equiv ( 2 \sinsthw )^2
\ , \quad \quad
  c \equiv ( 1 - \sinsthw )^2
\quad ,
\label{2p11}
\eeq
where
the squared weak-mixing angle is
\beq
  \sinsthw \approx 0.2325
\quad .
\label{2p12}
\eeq

The $A$ coefficients are given by
$$
 A_{\nueb} = A_{\nue} \equiv \Aek \ergk
$$
\beq
    A_{\nutb} = A_{\nut} = A_{\numb} = A_{\num}
      \equiv \Amk \ergk
\quad .
\label{2p9}
\eeq

The $B$ coefficients are
$$
   B_{\nueb} = B_{\nue} \equiv \Bek \ergkfac
$$
\beq
   B_{\nutb} = B_{\nut} = B_{\numb} = B_{\num}
    \equiv \Bmk \ergkfac
\quad .
\label{2p10}
\eeq

The $C$ coefficient
are expressed in terms of functions $\bar g_i$,
presented in Eq.\ \bref{ap8} below.
For the electron neutrino, we find
$$
  \CB_{\nue \nue }   =
       \Ceko \tgone  + \Cekt \tgthr   +
     2 \Ceko \tgtwo  +     3 \tgzero
\quad ,
$$
$$
  \CB_{\nue \num }   =
             \tgone  +     2 \tgthr   +
           2 \tgtwo  +     3 \tgzero
\quad ,
\quad\quad
  \CB_{\nue \nut }   =
             \tgone  +     2 \tgthr   +
           2 \tgtwo  +     3 \tgzero
\quad ,
$$
\beq
  \CB_{\nue \nueb }  =
           4 \tgone  + \Ceko \tgzero
\quad ,
\quad\quad
  \CB_{\nue \numb }  =
           2 \tgone  +       \tgzero
\quad ,
\quad\quad
  \CB_{\nue \nutb }  =
           2 \tgone  +       \tgzero
\quad .
\label{ap1}
\eeq
For the muon neutrino, we find
$$
  \CB_{\num \num }   =
       \Cmko \tgone  + \Cmkt \tgthr   +
     2 \Cmko \tgtwo  +     3 \tgzero
\quad ,
$$
$$
  \CB_{\num \nue }   =
             \tgone  +     2 \tgthr   +
           2 \tgtwo  +     3 \tgzero
\quad ,
\quad\quad
  \CB_{\num \nut }   =
             \tgone  +     2 \tgthr   +
           2 \tgtwo  +     3 \tgzero
\quad ,
$$
\beq
  \CB_{\num \nueb}   =
           2 \tgone  +       \tgzero
\quad ,
\quad\quad
  \CB_{\num \numb}   =
           4 \tgone  + \Cmko \tgzero
\quad ,
\quad\quad
  \CB_{\num \nutb}   =
           2 \tgone  +       \tgzero
\quad .
\label{ap2}
\eeq
For the tau neutrino, we find
$$
  \CB_{\nut \nut }   =
       \Cmko \tgone  + \Cmkt \tgthr   +
     2 \Cmko \tgtwo  +     3 \tgzero
\quad ,
$$
$$
  \CB_{\nut \nue }   =
             \tgone  +     2 \tgthr   +
           2 \tgtwo  +     3 \tgzero
\quad ,
\quad\quad
  \CB_{\nut \num }   =
             \tgone  +     2 \tgthr   +
           2 \tgtwo  +     3 \tgzero
\quad ,
$$
\beq
  \CB_{\nut \nueb}   =
           2 \tgone  +       \tgzero
\quad ,
\quad\quad
  \CB_{\nut \numb}   =
           2 \tgone  +       \tgzero
\quad ,
\quad\quad
  \CB_{\nut \nutb}   =
           4 \tgone  + \Cmko \tgzero
\quad .
\label{ap3}
\eeq
For the electron antineutrino, we find
$$
  \CB_{\nueb \nue }  =
           4 \tgone  + \Ceko \tgzero
\quad ,
\quad\quad
  \CB_{\nueb \num }  =
           2 \tgone  +       \tgzero
\quad ,
\quad\quad
  \CB_{\nueb \nut }  =
           2 \tgone  +       \tgzero
\quad ,
$$
$$
  \CB_{\nueb \nueb}  =
       \Ceko \tgone  + \Cekt \tgthr   +
     2 \Ceko \tgtwo  +     3 \tgzero
\quad ,
$$
\beq
  \CB_{\nueb \numb}  =
             \tgone  +     2 \tgthr   +
           2 \tgtwo  +     3 \tgzero
\quad ,
\quad\quad
  \CB_{\nueb \nutb}  =
             \tgone  +      2 \tgthr   +
           2 \tgtwo  +      3 \tgzero
\quad .
\label{ap4}
\eeq
For the muon antineutrino, we find
$$
  \CB_{\numb \nue }   =
            2 \tgone  +       \tgzero
\quad ,
\quad\quad
  \CB_{\numb \num }   =
            4 \tgone  + \Cmko \tgzero
\quad ,
\quad\quad
  \CB_{\numb \nut }   =
            2 \tgone  +       \tgzero
\quad ,
$$
$$
  \CB_{\numb \numb}   =
        \Cmko \tgone  + \Cmkt \tgthr   +
      2 \Cmko \tgtwo  +     3 \tgzero
\quad ,
$$
\beq
  \CB_{\numb \nueb}   =
              \tgone  +     2 \tgthr   +
            2 \tgtwo  +     3 \tgzero
\quad ,
\quad\quad
  \CB_{\numb \nutb}   =
              \tgone  +     2 \tgthr   +
            2 \tgtwo  +     3 \tgzero
\quad .
\label{ap5}
\eeq
For the tau antineutrino, we find
$$
  \CB_{\nutb \nue }   =
            2 \tgone  +       \tgzero
\quad ,
\quad\quad
  \CB_{\nutb \num }   =
            2 \tgone  +       \tgzero
\quad ,
\quad\quad
  \CB_{\nutb \nut }   =
            4 \tgone  + \Cmko \tgzero
\quad ,
$$
$$
  \CB_{\nutb \nutb}   =
        \Cmko \tgone  + \Cmkt \tgthr   +
      2 \Cmko \tgtwo  +     3 \tgzero
\quad ,
$$
\beq
  \CB_{\nutb \nueb}   =
              \tgone  +     2 \tgthr   +
            2 \tgtwo  +     3 \tgzero
\quad ,
\quad\quad
  \CB_{\nutb \numb}   =
              \tgone  +     2 \tgthr   +
            2 \tgtwo  +     3 \tgzero
\quad .
\label{ap6}
\eeq

In Eqs.\ \bref{ap1}--\bref{ap6},
we use the abbreviations
\beq
\Ceko \equiv a+b+6
\ , \quad
\Cmko \equiv b+c+6
\ , \quad
\Cekt \equiv 2 (a+b)+10
\ , \quad
\Cmkt \equiv 2 (b+c)+10
\quad .
\label{ap7}
\eeq
The $\bar g$ are scaled versions of the functions $g$
of ref.\ \ct{dt92a}:
$$
 \tgzero \EoTEpoT \equiv
   - \fr{1}{18} \fr{E}{T}
   \left( \fr{E'}{T} \right)^3 \fexp
\quad ,
$$
$$
  \tgone \EoTEpoT \equiv  \ptemp g_1 \EoTEpoT
\quad ,
$$
$$
  \tgtwo \EoTEpoT \equiv  \ptemp g_2 \EoTEpoT
\quad ,
$$
\beq
  \tgthr \EoTEpoT \equiv  \ptemp g_3 \EoTEpoT
\quad .
\label{ap8}
\eeq
For completeness we give the unbarred $g_i \EoTEpoT $:
$$
  g_1 \EoTEpoT =
  \int_{ \left| v \right| }^w {dy}
  e^{-{y \over 2}}\left( {v^2-y^2} \right)^2
\quad ,
$$
$$
  g_2 \EoTEpoT =
   -\int_{ \left| v \right| }^w {{{dy} \over {2y^2}}}
   e^{-{y \over 2}}\left( {v^2-y^2} \right)^2
   \left( {2w + wy + y^2} \right)
\quad ,
$$
\bea
  g_3 \EoTEpoT &=&
   \int_{ \left| v \right| }^w {{{dy} \over {4y^4}}}
   e^{-{y \over 2}}\left( {v^2-y^2} \right)^2
  \nonumber \\
  && \times \left( { 12w^2 + 6w^2y
   + \left( {w^2+4w-4} \right)y^2 +
   2\left( {w-1} \right)y^3 + y^4 } \right)
\quad ,
\label{ap9}
\eea
where the abbreviations
\beq
  v \equiv {{\left( {E-E'} \right)} \over T}
\ , \quad \quad
  w\equiv {{\left( {E+E'} \right)} \over T}
\ , \quad \quad
  x\equiv {E \over T}
\quad
\label{ap10}
\eeq
are used.
The lower limit of each integration
in Eq.\ \bref{ap9}
is the absolute value of $v$.
Note that the above expression for $g_3$ differs from
that in ref.\ \ct{dt92a},
which is incorrectly typeset.

\bigskip
\vglue 0.2cm
{\bf\large\noindent
Appendix B: Coefficients
for Combined Production-Oscillation}
\vglue 0.2cm
\setcounter{section}{12}
\setcounter{equation}{0}
\def\theequation{B.\arabic{equation}}

Here, we express the coefficients
$A_{ab}$, $A_{\bar a\bar b}$,
$B_a$, $B_{\bar a}$,
$C_{a b}$, $C_{a \bar b}$, $C_{\bar a b}$, and $C_{\bar a \bar b}$
in the vector basis in terms of the
coefficients
$A_{\nu_f}$,
$\tilde B_{\nu_f}$,
and
$\tilde C_{\nu_f \nu_{f'}}$.
Since all coefficients
are functions of the scaled initial and final neutrino energies
$\frac{E}{T}$ and $\frac{E'}{T}$,
we drop this dependence everywhere to save space.

The nonzero $A_{ab}$ and $A_{\bar a\bar b}$ are
\beq
  A_{00} = A_{11} =
    {1 \over 2}\left( {A_{\nu_e} + A_{\nu_\mu }} \right)
\quad , \quad \quad
  A_{01} = A_{10} = {1 \over 2}
               \left( {A_{\nu_e} - A_{\nu_\mu }} \right)
\quad ,
\quad
  A_{44} = A_{\nu_\tau }
\quad
\label{5p7}
\eeq
and
\beq
  A_{\bar 0 \bar 0} = A_{\bar 1 \bar 1} =
    {1 \over 2}
  \left( {A_{\bar \nu_e} + A_{\bar \nu_\mu }} \right)
\quad , \quad \quad
  A_{\bar 0 \bar 1} = A_{\bar 1 \bar 0} =
    {1 \over 2}
  \left( {A_{\bar \nu_e} - A_{\bar \nu_\mu }} \right)
\quad ,
\quad
  A_{\bar 4 \bar 4} = A_{\bar \nu_\tau }
\quad .
\label{5p10}
\eeq

The nonzero $B_a$ and $B_{\bar a}$ are
\beq
  B_0 =
        \tilde B_{\nu_e} + \tilde B_{\nu_\mu }
\quad , \quad \quad
  B_1 =
        \tilde B_{\nu_e} - \tilde B_{\nu_\mu }
\quad , \quad \quad
  B_4 = \tilde B_{\nu_\tau }
\quad .
\label{5p8}
\eeq
and
\beq
  B_{\bar 0} =
    \tilde B_{\bar \nu_e} + \tilde B_{\bar \nu_\mu }
\quad , \quad \quad
  B_{\bar 1} =
    \tilde B_{\bar \nu_e} - \tilde B_{\bar \nu_\mu }
\quad , \quad \quad
  B_{\bar 4} = \tilde B_{\bar \nu_\tau }
\quad .
\label{5p11}
\eeq

The remaining nonzero coefficients for neutrinos are
$$
  C_{00} =
  {1 \over 2}\left( {
   \tilde C_{\nu_e   \nu_e}    +
   \tilde C_{\nu_\mu \nu_e}    +
   \tilde C_{\nu_e   \nu_\mu } +
   \tilde C_{\nu_\mu \nu_\mu }
                     } \right)
\quad ,
$$
$$
  C_{01} =
  {1 \over 2}\left( {
   \tilde C_{\nu_e   \nu_e}    +
   \tilde C_{\nu_\mu \nu_e}    -
   \tilde C_{\nu_e   \nu_\mu } -
   \tilde C_{\nu_\mu \nu_\mu }
                     } \right)
\quad ,
$$
$$
  C_{10} =
  {1 \over 2}\left( {
   \tilde C_{\nu_e   \nu_e}    -
   \tilde C_{\nu_\mu \nu_e}    +
   \tilde C_{\nu_e   \nu_\mu } -
   \tilde C_{\nu_\mu \nu_\mu }
                     } \right)
\quad ,
$$
$$
  C_{11} =
  {1 \over 2}\left( {
   \tilde C_{\nu_e   \nu_e}    -
   \tilde C_{\nu_\mu \nu_e}    -
   \tilde C_{\nu_e   \nu_\mu } +
   \tilde C_{\nu_\mu \nu_\mu }
                     } \right)
\quad ,
$$
$$
  C_{04} =
              \left( {
   \tilde C_{\nu_e   \nu_\tau} +
   \tilde C_{\nu_\mu \nu_\tau}
                     } \right)
\quad ,
\quad\quad
  C_{14} =
              \left( {
   \tilde C_{\nu_e   \nu_\tau} -
   \tilde C_{\nu_\mu \nu_\tau}
                     } \right)
\quad ,
$$
$$
  C_{40} =
   {1 \over 2}\left( {
   \tilde C_{\nu_e   \nu_\tau} +
   \tilde C_{\nu_\mu \nu_\tau}
                     } \right)
\quad ,
\quad\quad
  C_{41} =
   {1 \over 2}\left( {
   \tilde C_{\nu_e   \nu_\tau} -
   \tilde C_{\nu_\mu \nu_\tau}
                     } \right)
\quad ,
\quad\quad
  C_{44} =
   \tilde C_{\nu_\tau   \nu_\tau}
\quad ,
$$
$$
  C_{0 \bar 0} =
  {1 \over 2}\left( {
   \tilde C_{\nu_e    \bar \nu_e}    +
   \tilde C_{\nu_\mu  \bar \nu_e}    +
   \tilde C_{\nu_e    \bar \nu_\mu } +
   \tilde C_{\nu_\mu  \bar \nu_\mu }
                     } \right)
\quad ,
$$
$$
  C_{0 \bar 1} =
  {1 \over 2}\left( {
   \tilde C_{\nu_e    \bar \nu_e}    +
   \tilde C_{\nu_\mu  \bar \nu_e}    -
   \tilde C_{\nu_e    \bar \nu_\mu } -
   \tilde C_{\nu_\mu  \bar \nu_\mu }
                     } \right)
\quad ,
$$
$$
  C_{1 \bar 0} =
  {1 \over 2}\left( {
   \tilde C_{\nu_e    \bar \nu_e}    -
   \tilde C_{\nu_\mu  \bar \nu_e}    +
   \tilde C_{\nu_e    \bar \nu_\mu } -
   \tilde C_{\nu_\mu  \bar \nu_\mu }
                     } \right)
\quad ,
$$
$$
  C_{1 \bar 1} =
  {1 \over 2}\left( {
   \tilde C_{\nu_e    \bar \nu_e}    -
   \tilde C_{\nu_\mu  \bar \nu_e}    -
   \tilde C_{\nu_e    \bar \nu_\mu } +
   \tilde C_{\nu_\mu  \bar \nu_\mu }
                     } \right)
\quad ,
$$
$$
  C_{0 \bar 4} =
              \left( {
   \tilde C_{\nu_e    \bar \nu_\tau} +
   \tilde C_{\nu_\mu  \bar \nu_\tau}
                     } \right)
\quad ,
\quad\quad
  C_{1 \bar 4} =
              \left( {
   \tilde C_{\nu_e    \bar \nu_\tau} -
   \tilde C_{\nu_\mu  \bar \nu_\tau}
                     } \right)
\quad ,
$$
\beq
  C_{4 \bar 0} =
   {1 \over 2}\left( {
   \tilde C_{\nu_e    \bar \nu_\tau} +
   \tilde C_{\nu_\mu  \bar \nu_\tau}
                     } \right)
\quad ,
\quad\quad
  C_{4 \bar 1} =
   {1 \over 2}\left( {
   \tilde C_{\nu_e    \bar \nu_\tau} -
   \tilde C_{\nu_\mu  \bar \nu_\tau}
                     } \right)
\quad ,
\quad\quad
  C_{4 \bar 4} =
   \tilde C_{\nu_\tau    \bar \nu_\tau}
\quad .
\label{bp1}
\eeq

For antineutrinos,
the remaining nonzero coefficients are
$$
  C_{ \bar 00} =
  {1 \over 2}\left( {
   \tilde C_{ \bar \nu_e   \nu_e}    +
   \tilde C_{ \bar \nu_\mu \nu_e}    +
   \tilde C_{ \bar \nu_e   \nu_\mu } +
   \tilde C_{ \bar \nu_\mu \nu_\mu }
                     } \right)
\quad ,
$$
$$
  C_{ \bar 01} =
  {1 \over 2}\left( {
   \tilde C_{ \bar \nu_e   \nu_e}    +
   \tilde C_{ \bar \nu_\mu \nu_e}    -
   \tilde C_{ \bar \nu_e   \nu_\mu } -
   \tilde C_{ \bar \nu_\mu \nu_\mu }
                     } \right)
\quad ,
$$
$$
  C_{ \bar 10} =
  {1 \over 2}\left( {
   \tilde C_{ \bar \nu_e   \nu_e}    -
   \tilde C_{ \bar \nu_\mu \nu_e}    +
   \tilde C_{ \bar \nu_e   \nu_\mu } -
   \tilde C_{ \bar \nu_\mu \nu_\mu }
                     } \right)
\quad ,
$$
$$
  C_{ \bar 11} =
  {1 \over 2}\left( {
   \tilde C_{ \bar \nu_e   \nu_e}    -
   \tilde C_{ \bar \nu_\mu \nu_e}    -
   \tilde C_{ \bar \nu_e   \nu_\mu } +
   \tilde C_{ \bar \nu_\mu \nu_\mu }
                     } \right)
\quad ,
$$
$$
  C_{ \bar 04} =
              \left( {
   \tilde C_{ \bar \nu_e   \nu_\tau} +
   \tilde C_{ \bar \nu_\mu \nu_\tau}
                     } \right)
\quad ,
\quad\quad ,
  C_{ \bar 14} =
              \left( {
   \tilde C_{ \bar \nu_e   \nu_\tau} -
   \tilde C_{ \bar \nu_\mu \nu_\tau}
                     } \right)
\quad ,
$$
$$
  C_{ \bar 40} =
   {1 \over 2}\left( {
   \tilde C_{ \bar \nu_e   \nu_\tau} +
   \tilde C_{ \bar \nu_\mu \nu_\tau}
                     } \right)
\quad ,
\quad\quad ,
  C_{ \bar 41} =
   {1 \over 2}\left( {
   \tilde C_{ \bar \nu_e   \nu_\tau} -
   \tilde C_{ \bar \nu_\mu \nu_\tau}
                     } \right)
\quad ,
\quad\quad ,
  C_{ \bar 44} =
   \tilde C_{ \bar \nu_\tau   \nu_\tau}
\quad ,
$$
$$
  C_{ \bar 0 \bar 0} =
  {1 \over 2}\left( {
   \tilde C_{ \bar \nu_e    \bar \nu_e}    +
   \tilde C_{ \bar \nu_\mu  \bar \nu_e}    +
   \tilde C_{ \bar \nu_e    \bar \nu_\mu } +
   \tilde C_{ \bar \nu_\mu  \bar \nu_\mu }
                     } \right)
\quad ,
$$
$$
  C_{ \bar 0 \bar 1} =
  {1 \over 2}\left( {
   \tilde C_{\ \bar nu_e    \bar \nu_e}    +
   \tilde C_{ \bar \nu_\mu  \bar \nu_e}    -
   \tilde C_{ \bar \nu_e    \bar \nu_\mu } -
   \tilde C_{ \bar \nu_\mu  \bar \nu_\mu }
                     } \right)
\quad ,
$$
$$
  C_{ \bar 1 \bar 0} =
  {1 \over 2}\left( {
   \tilde C_{ \bar \nu_e    \bar \nu_e}    -
   \tilde C_{ \bar \nu_\mu  \bar \nu_e}    +
   \tilde C_{ \bar \nu_e    \bar \nu_\mu } -
   \tilde C_{ \bar \nu_\mu  \bar \nu_\mu }
                     } \right)
\quad ,
$$
$$
  C_{ \bar 1 \bar 1} =
  {1 \over 2}\left( {
   \tilde C_{ \bar \nu_e    \bar \nu_e}    -
   \tilde C_{ \bar \nu_\mu  \bar \nu_e}    -
   \tilde C_{ \bar \nu_e    \bar \nu_\mu } +
   \tilde C_{ \bar \nu_\mu  \bar \nu_\mu }
                     } \right)
\quad ,
$$
$$
  C_{ \bar 0 \bar 4} =
              \left( {
   \tilde C_{ \bar \nu_e    \bar \nu_\tau} +
   \tilde C_{ \bar \nu_\mu  \bar \nu_\tau}
                     } \right)
\quad ,
\quad\quad ,
  C_{ \bar 1 \bar 4} =
              \left( {
   \tilde C_{ \bar \nu_e    \bar \nu_\tau} -
   \tilde C_{ \bar \nu_\mu  \bar \nu_\tau}
                     } \right)
\quad ,
$$
\beq
  C_{ \bar 4 \bar 0} =
   {1 \over 2}\left( {
   \tilde C_{ \bar \nu_e    \bar \nu_\tau} +
   \tilde C_{ \bar \nu_\mu  \bar \nu_\tau}
                     } \right)
\quad ,
\quad\quad ,
  C_{ \bar 4 \bar 1} =
   {1 \over 2}\left( {
   \tilde C_{ \bar \nu_e    \bar \nu_\tau} -
   \tilde C_{ \bar \nu_\mu  \bar \nu_\tau}
                     } \right)
\quad ,
\quad\quad ,
  C_{ \bar 4 \bar 4} =
   \tilde C_{ \bar \nu_\tau    \bar \nu_\tau}
\quad .
\label{bp2}
\eeq

\vglue 0.6cm
{\bf\large\noindent References}
\vglue 0.4cm

\def\plb #1 #2 #3 {Phys.\ Lett.\ B #1 (19#2) #3.}
\def\mpl #1 #2 #3 {Mod.\ Phys.\ Lett.\ A #1 (19#2) #3.}
\def\prl #1 #2 #3 {Phys.\ Rev.\ Lett.\ #1 (19#2) #3.}
\def\pr #1 #2 #3 {Phys.\ Rev.\ #1 (19#2) #3.}
\def\prd #1 #2 #3 {Phys.\ Rev.\ D #1 (19#2) #3.}
\def\npb #1 #2 #3 {Nucl.\ Phys.\ B#1 (19#2) #3.}
\def\ptp #1 #2 #3 {Prog.\ Theor.\ Phys.\ #1 (19#2) #3.}
\def\jmp #1 #2 #3 {J.\ Math.\ Phys.\ #1 (19#2) #3.}
\def\nat #1 #2 #3 {Nature #1 (19#2) #3.}
\def\prs #1 #2 #3 {Proc.\ Roy.\ Soc.\ (Lon.) A #1 (19#2) #3.}
\def\ajp #1 #2 #3 {Am.\ J.\ Phys.\ #1 (19#2) #3.}
\def\lnc #1 #2 #3 {Lett.\ Nuov.\ Cim. #1 (19#2) #3.}
\def\nc #1 #2 #3 {Nuov.\ Cim.\ A#1 (19#2) #3.}
\def\jpsj #1 #2 #3 {J.\ Phys.\ Soc.\ Japan #1 (19#2) #3.}
\def\ant #1 #2 #3 {At. Dat. Nucl. Dat. Tables #1 (19#2) #3.}
\def\nim #1 #2 #3 {Nucl.\ Instr.\ Meth.\ B#1 (19#2) #3.}

\newpage
\noindent
FIGURE CAPTIONS

\bigskip
\noindent
Figure 1: Pure-production profiles.
For zero mixing angle and at $T = 0.75$ MeV,
the excess-neutrino number per unit energy
in the canonical comoving volume is shown
as a function of energy.
The solid line is electron neutrinos,
the dashed line is muon neutrinos,
the dotted line is their difference,
and the dashed-dotted line is their sum.

\bigskip
\noindent
Figure 2: Comparison of pure-production profiles.
For zero mixing angle and at $T = 0.75$ MeV,
the excess-neutrino number per unit energy
in the canonical comoving volume is shown
as a function of energy.
The solid line is our electron-neutrino results,
the dotted line is the difference
between our electron- and muon-neutrino results,
and the dashed line is the analytical approximation
of ref.\ \ct{df}.

\bigskip
\noindent
Figure 3: Time development of nonequilibrium distortions.
The solid line is the total excess of electron neutrinos
in the canonical comoving volume as a function of time.
The dashed line is the total muon-neutrino excess,
the dotted line is the difference between electron and
muon neutrinos,
and the dashed-dotted line is their sum.

\bigskip
\noindent
Figure 4: Production profiles for $\Delta = 10^{-6}$ eV$^2$
and $\sin^2 2 \theta = 0.81$.
The excess-neutrino number per unit energy
in the canonical comoving volume is shown
as a function of energy.
The solid line is electron neutrinos,
the dashed line is muon neutrinos,
the dotted line is their difference,
the dashed-dotted line is their sum,
and the short-dashed line is tau neutrinos.

\bigskip
\noindent
Figure 5: Analytical approximations to profile
plots for $\Delta = 10^{-6}$ eV$^2$, $\ss=0.81$.
The solid lines are the data,
the dashed lines are the projection approximation
\bref{8p3},
and the dotted lines are
the rotation approximation \bref{8p2}.
(a) The first component of the neutrino vector
as a function of energy.
(b) The second component of the neutrino vector
as a function of energy.

\bigskip
\noindent
Figure 6: Components of the average
neutrino vector as a function of time for the case with
$\Delta = 10^{-6}$ eV$^2$ and $\ss=0.81$.
The dashed-dotted line
is the zeroth component,
the dotted line
is the first component,
the solid line is the second component,
and the short-dashed line
is the fourth component.
The third component is indistinguishable from zero
on the scale of the figure.

\bigskip
\noindent
Figure 7: Components of the normalized average
neutrino vector $\vec r_v(t) $
as a function of time for the case with
$\Delta = 10^{-12}$ eV$^2$ and $\ss = 0.81$.
The solid lines represent the new data,
while the dashed lines represent data
from earlier work \ct{ks94a}.
The curves above the horizonal axis
are the component $r_{v1}$,
while the curves below it are the component $r_{v2}$.

\vfill\eject

\begin{thebibliography}{xx}

\def\NPB#1#2#3{ {Nucl.{\,}Phys.{\,}}{\bf B{#1}} ({#3}) {#2}}
\def\PLB#1#2#3{ {Phys.{\,}Lett.{\,}}{\bf B{#1}} ({#3}) {#2}}
\def\PRL#1#2#3{ {Phys.{\,}Rev.{\,}Lett.{\,}}
  {\bf  {#1}} ({#3}) {#2}}
\def\PRD#1#2#3{ {Phys.{\,}Rev.{\,}}{\bf D{#1}} ({#3}) {#2}}

\bibitem{kt90a}
See, for example,
E.{\,}W.{\,}Kolb and M.{\,}S.{\,}Turner,
\it The Early Universe \rm
(Addison Wesley, Redwood City, CA 1990).

\bibitem{bv92}
F.\ Boehm and P.\ Vogel,
\it The Physics of Massive Neutrinos, 2nd ed.\ \rm
(Cambridge University Press, Cambridge, 1992).

\bibitem{d81a}
A.\ D.\ Dolgov,
Sov.\ J.\ Nucl.\ Phys.\ {\bf 33} (1981) 700.

\bibitem{s87a}
L.\ Stodolsky,
\PRD{36}{2273}{1987}.

\bibitem{l}
P.{\,}Langacker, S.{\,}T.{\,}Petcov,
G.{\,}Steigman, and S.{\,}Toshev, \\
\NPB{282}{589}{1987}.

\bibitem{n}
D.\ Notzold and G.\ Raffelt,
\NPB{307}{924}{1988}.

\bibitem{s}
M.{\,}J.{\,}Savage, R.{\,}A.{\,}Malaney
and G.{\,}M.{\,}Fuller,
Astrophys.{\,}J.{\,}{\bf 368} (1991) 1.

\bibitem{ekm91a}
K.{\,}Enqvist, K.{\,}Kainulainen and J.{\,}Maalampi,
\NPB{349}{754}{1991}.

\bibitem{ekt92}
K.{\,}Enqvist, K.{\,}Kainulainen and M.{\,}Thomson,
\NPB{373}{498}{1992}.

\bibitem{ssf93a}
X.\ Shi, D.\ N.\ Schramm and B.\ D.\ Fields,
\PRD{48}{2563}{1993}.

\bibitem{kps93a}
V.{\,}A.{\,}Kosteleck\'y,
J.{\,}Pantaleone and S.{\,}Samuel,
\PLB{315}{46}{1993}.

\bibitem{ks93a}
V.{\,}A.{\,}Kosteleck\'y and S.{\,}Samuel,
\PLB{318}{127}{1993}.

\bibitem{ks94a}
V.{\,}A.{\,}Kosteleck\'y and S.{\,}Samuel,
\PRD{49}{1740}{1994}.

\bibitem{mt94a}
B.\ H.\ J.\ McKellar and M.\ J.\ Thomson
\PRD{49}{2710}{1994}.

\bibitem{dt92a}
S.{\,}Dodelson and M.{\,}S.{\,}Turner,
\PRD{46}{3372}{1992}; \\
B.{\,}Fields, S.{\,}Dodelson and M.{\,}S.{\,}Turner,
\PRD{47}{4309}{1993}.

\bibitem{df}
A.{\,}D.{\,}Dolgov and M.{\,}Fukugita,
\PRD{46}{5378}{1992}.

\bibitem{pantaleone}
J.{\,}Pantaleone,
\PLB{287}{128}{1992},
\PRD{46}{510}{1992}.

\bibitem{sr93a}
G.{\,}Sigl and G.{\,}Raffelt,
\NPB{406}{423}{1993}.

\bibitem{samuel93a}
S.{\,}Samuel,
\PRD{48}{1462}{1993}.

\bibitem{msw}
L.{\,}Wolfenstein,
\PRD{17}{2369}{1978}; \\
\PRD{20}{2634}{1979}; \\
S.{\,}P.{\,}Mikheyev and A.{\,}Yu.{\,}Smirnov,
Yad.{\,}Fiz.{\,}{\bf 42} (1985) 1441; \\
Usp.{\,}Fiz.{\,}Nauk.{\,}{\bf 153} (1987) 3.

\bibitem{ks94b}
V.{\,}A.{\,}Kosteleck\'y and S.{\,}Samuel,
{\it Self-Maintained Coherent Oscillations
in Dense Neutrino Gases},
IUHET 291, MPI-PhT/94-80, CCNY-HEP-94/8 (December, 1994).

\end{thebibliography}
\end{document}